\documentclass[twocolumn,lineno]{aastex7}

\usepackage{color}
\usepackage{booktabs}
\usepackage{longtable}
\usepackage{soul}
\usepackage{bigstrut}
\usepackage{graphicx}
\usepackage{amsmath}

\usepackage{mathrsfs}
\usepackage{appendix}
\usepackage{tablefootnote}
\usepackage{tikz}
\usepackage{subcaption}

\usepackage{hyperref}
\usepackage{nameref}

\definecolor{DarkGreen}{RGB}{32,140,34}

\let\listofchanges\undefined
\usepackage[deletedmarkup=xout,commentmarkup=uwave,commandnameprefix=ifneeded]{changes}

\definechangesauthor[name=Max, color=teal]{MI}
\definechangesauthor[name=Will, color=orange]{WF}
\definechangesauthor[name=Chengcheng, color=magenta]{CX}
\definechangesauthor[name=Zoltan, color=DarkGreen]{ZH}

\defcitealias{Xin2021}{XH21}
\defcitealias{Xin2024}{XH24}

\newcommand{\Msun}{M_\odot}
\newcommand{\Mc}{\mathcal{M}}
\newcommand{\Mtot}{M_{\rm bin}}
\newcommand{\mi}{m_i}
\newcommand{\McDet}{(1+z)\Mc}
\newcommand{\wmid}{w_0}
\newcommand{\wdotmid}{\dot w_0}

\graphicspath{{./}{figures/}}

\revised{\today}

\shorttitle{Compact SMBHBs in LSST}
\shortauthors{Xin et al.}

\begin{document}

\title{Identifying Compact Chirping SMBHBs in LSST using Bayesian Analysis}

\correspondingauthor{Chengcheng Xin}

\author[0000-0003-3106-8182]{Chengcheng Xin}
\affil{Columbia University, Department of Astronomy, 550 West 120th Street, New York, NY, 10027, USA}
\email[show]{cx2204@columbia.edu}

\author[0000-0001-8830-8672]{Maximiliano Isi}
\affil{Center for Computational Astrophysics, Flatiron Institute, New York, NY 10010, USA}
\email{misi@flatironinstitute.org}

\author[0000-0003-1540-8562]{Will M. Farr}
\affil{Center for Computational Astrophysics, Flatiron Institute, New York, NY 10010, USA}
\affil{Department of Physics and Astronomy, Stony Brook University, Stony Brook, NY 11794-3800, USA}
\email{wfarr@flatironinstitute.org}

\author{Zolt{\'{a}}n~Haiman}
\affil{Columbia University, Department of Astronomy, 550 West 120th Street, New York, NY, 10027, USA}
\affil{Columbia University, Department of Physics, 550 West 120th Street, New York, NY, 10027, USA}
\affil{Institute of Science and Technology Austria (ISTA), Am Campus 1, Klosterneuburg 3400, Austria}
\email{}

\begin{abstract}
The Legacy Survey of Space and Time (LSST) is expected to observe up to ${\sim}100$ million quasars in the next decade. In this work, we show that it is possible to use such data to measure the characteristic frequency evolution of a ``chirp'' induced by gravitational waves, which can serve as robust evidence for the presence of a compact supermassive black-hole binary.
Following the LSST specifications, we generate mock lightcurves consisting of (i) a post-Newtonian chirp produced by orbital motion through, e.g., relativistic Doppler boosting, (ii) a damped random walk representing intrinsic quasar variability, and (iii) Gaussian photometric errors, while assuming non-uniform observations with extended gaps over a period of 10 yr.
Through a fully-Bayesian analysis, we show that we can simultaneously measure the chirp and noise parameters with little degeneracy between the two. For chirp signals with an amplitude of $A = 0.5$ mag and a range of times to merger ($t_m = 15{-}10^4$ yr), we can typically measure a non-zero amplitude and positive frequency derivative with over $5\sigma$ credibility. For binaries with $t_m = 50$ yr, we achieve $3\sigma$ ($5\sigma$) confidence that the signal is chirping for $A \gtrsim 0.1$ ($A > 0.2$).
Our analysis can take as little as 35 s (and typically $<$ 10 min) to run, making it scalable to a large number of lightcurves.
This implies that LSST could, on its own, establish the presence of a compact supermassive black-hole binary, and thus discover gravitational wave sources detectable by LISA and by Pulsar Timing Arrays.
\end{abstract}

\section{Introduction}
\label{sec:intro}
The Legacy Survey of Space and Time (LSST) by the Vera C. Rubin Observatory is expected to observe tens to over a hundred million quasars during its 10-year lifespan, which will begin at the end of 2025 \citep[][hereafter \citetalias{Xin2021}]{Xin2021}. 
Assuming that most of these quasars are associated with
coalescing supermassive black hole binaries (SMBHBs) triggered by galaxy mergers \citep{Begelman+1980,KauffmannHaehnelt2000,AlexanderHickox2012,Goulding+2018}, we can expect that a small fraction 
are compact inspiraling SMBHBs driven by gravitational wave (GW) emission. Electromagnetic (EM) observations of these binaries are ideal for multi-messenger studies when combined with their GW emission, which is expected in the nano-Hz to milli-Hz band. This makes them prime targets for the upcoming Laser Interferometer Space Antenna \citep[LISA;][]{LISA2024} and pulsar timing arrays \citep[PTAs;][]{Agazie2023}. 
\citetalias{Xin2021} estimate that up to $\sim$150 of the GW-driven binaries in LSST can be ultra-compact ($1{-}2$ day periods), meaning that they can merge and be detected as mHz GWs within the LISA band $5{-}15$ years later. The lightcurves of compact binaries in LSST are expected to exhibit distinct periodicity and period evolution, namely an EM ``chirp" \citep{Haiman2017}. Detecting such a chirp signal can serve as smoking-gun evidence for a GW-driven SMBHB. This is a novel task that is expected to be challenging unless the GWs of the binary are also observed \citep[][hereafter XH24]{Xin2024}; \citetalias{Xin2024} show that LISA data can make it possible to identify the signal in archival LSST data.

Generally, as long as they are luminous, SMBHBs can be identified in time-domain surveys as quasars with periodic variability \citep{Haiman2009a}.
In the past decade, SMBHB candidates have been identified in time-domain surveys, including a total of $\sim$400 candidates identified in the Catalina Real-Time Transient survey \citep[CRTS;][]{Graham2015a}, the Zwicky Transient Facility \citep[CRTS+ZTF;][]{Foustoul2025}, the Palomar Transient Factory \citep[PTF;][]{Charisi2016}, the Sloan Digital Sky Surveys and Dark Energy Survey \citep[SDSS+DES;][]{Chen2020}, and the Gaia space observatory \citep{Huijse2025}. The identification of these candidates suffers from false-positive contamination due to quasar ``red noise" \citep{Vaughan2016} and observational limitations, including the cadence, photometric uncertainty, baseline and depth of the surveys \citep{Witt2022}. LSST will observe a large quasar catalog, including a never-before-seen population of faint AGNs (down to $i=26$ magnitude) with high cadence and a long baseline~\citep{LSSTScienceCollaboration2009}. 
The better ``data quality" of LSST should allow us to (i) identify rare, short-lived binaries with shorter periods ($P\sim$ months to days) that exhibit a chirp, and (ii) distinguish SMBHBs from stochastic quasar variability.

The Lomb-Scargle (LS) periodogram \citep{Lomb1976,Scargle1982} is commonly used in studies to detect periodicity in quasar lightcurves. However, the LS periodogram is suboptimal for identifying chirping signals and inferring multiple binary and noise parameters beyond the orbital period. Periodograms also struggle to quantify the significance of detected peaks and to identify significant rates of change in frequency.\footnote{\cite{Robnik2024} used a ``null-signal template" as an alternative approach to more efficiently test periodic signals.} \citetalias{Xin2024} have shown that the LS periodogram appears noisy, and any significant peak vanishes if a chirp component is present.
Some past works have employed Bayesian methods to identify periodic SMBHBs in existing time-domain data \citep[e.g.,][]{DOrazio2015,Huijse2025,Foustoul2025} and in mock LSST data \citep{Witt2022}. However, they do not address our question: whether chirp and noise properties, in additional to periodicity, are measurable---and for which binaries are they measurable---within the real LSST quasar population as predicted by the quasar luminosity function. 

Our goal is to identify chirping SMBHBs in LSST quasar lightcurves and to assess the detectability of various properties of these binaries.
We develop a fully Bayesian framework to analyze LSST data and to extract the properties of the EM chirp, if present.
We model the lightcurves as composed of (i) quasi-sinusoidal variability induced by relativistic Doppler boost (DB), with GW-driven frequency evolution evaluated to leading post-Newtonian order, (ii) a ``damped random walk" (DRW) representing stochastic variability intrinsic to the quasar, and (iii) Gaussian photometric errors; we also allow for both non-uniform observing cadences and extended gaps in observations.
An efficient likelihood parameterization and the use of a gradient-assisted, hardware-accelerated sampler allow us to efficiently analyze a large number of lightcurves, and show that LSST will be able to measure the chirp in sufficiently compact SMBHB sources.

We find that chirp and noise parameters are measurable for a wide range of compact SMBHBs with $M=10^5{-}10^9{\rm M_{\odot}}$, $z=0{-}6$ and $t_m=15{-}10^4$ yr, assuming the chirp amplitude ($A$) is 0.5 mag, with periods $2{-}300$ d. 
The statistical confidence depends on $A$ and other system parameters. We perform our analysis for fiducial binaries ($t_m=50$ yr), and we find that can achieve at least $3\sigma$ confidence in measuring binary parameters when $A>0.1$. The sections in this paper are organized as follows.
In \S\ref{sec:method}, we discuss our selection of simulated LSST source parameters (\S\ref{subsec:method-sampling}), as well as our method to simulate mock lightcurves (\S\ref{subsec:method-lc}) and to analyze them (\S\ref{subsec:method-bayes})
We report our results in \S\ref{sec:results}.
Finally, in \S\ref{sec:discussion}, we summarize our findings and discuss future plans, including expanding this study beyond the DB and DRW assumptions.

\section{Methods} \label{sec:method}

\subsection{Sampling LSST binary quasars} \label{subsec:method-sampling}

\begin{figure}
    \centering
    \includegraphics[width=\columnwidth]{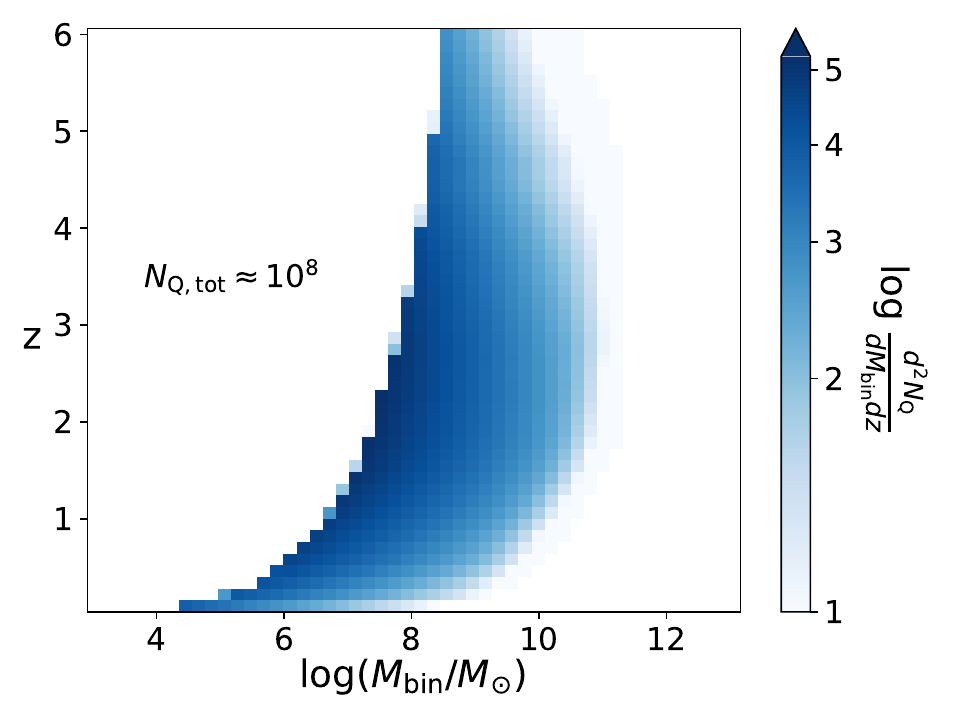}
    \caption{Expected number density of binary quasars in LSST (color) as a function of total binary mass (abscissa) and redshift (ordinate), using the extrapolated quasar luminosity function from \cite{Kulkarni2019} and assuming that the LSST co-added $i$-band magnitude threshold is $m_i=26$ (left edge of the contour). The total number of expected LSST quasars is $\sim$120 million. }
    \label{fig:n_lsst}
\end{figure}

To define a potential set of LSST sources, we follow \citetalias{Xin2021} to estimate the number of quasars as a function of their brightness and redshift. Using the extrapolated quasar luminosity function \citep{Kulkarni2019}, we estimate that LSST is expected to observe up to $N_{\rm total}\sim 100$ million quasars during its 10-year survey, assuming that the detection limit reaches magnitude $m_i =$ 26 in the $i$-band~\citep{LSSTScienceCollaboration2009}. The binary mass ($\Mtot$) can be estimated using $m_i$; assuming a typical quasar spectral energy distribution \citep[][see also \citetalias{Xin2024}, Eq.~10]{Haiman2009a} and a fiducial Eddington ratio $f_{\rm Edd}=0.3$, then
\begin{align} \label{eq:mi}
        m_i = 26 + 2.5 \log &\left[ \left( \frac{ f_{\rm Edd}}{0.3}\right)^{-1} \left( \frac{\Mtot}{3\times 10^7 \Msun}\right)^{-1} \right. \nonumber \\
        &\left.\times \left( \frac{d_L(z)}{d_L(z=2)}\right)^2 \right] ,
\end{align}
where $d_L(z)$ is the luminosity distance as a function of redshift $z$ in the $\Lambda-$CDM cosmology we adopted with $H_0=70$ km/s/Mpc, $\Omega_m=0.3$ and $\Omega_{\Lambda}=0.7$.
In Figure~\ref{fig:n_lsst}, we show the distribution of the number of quasars $N_Q$ as a function of $\Mtot$ and $z$.
The quasar density peaks around $z\approx 2$ and $\Mtot\approx 10^5 \Msun$, consistent with the measured quasar luminosity function. The left edge of the contour represents the LSST magnitude limit of $m_i=26$, per the relationship in Eq.~\eqref{eq:mi}. 
We use Fig.~\ref{fig:n_lsst} to draw masses and redshifts of putative LSST sources.

To simulate an LSST observation, we further need to specify the orbital separation or, equivalently (for a fixed mass ratio $q\equiv m_2/m_1\leq 1$; see below), the remaining time to merger.
By a simple counting argument, the abundance of binaries with (detector-frame) time to merger $t_m$ must be proportional to $t_m/t_Q$, where we take $t_Q = 10^7\, {\rm yr}$ to be the typical observable lifetime of a luminous quasar \citep{Martini+2004}.
For instance, given the results in Fig.~\ref{fig:n_lsst}, the total number of binaries with $t_m=50 \, {\rm yr}$ is approximately $t_m / t_{Q} \times f_{\rm bin} \times N_{\rm Q, tot} = (50 \,{\rm yr}/10^7 \, {\rm yr}) \times 10^8 \approx 500$, assuming that the fraction of quasars associated with SMBHBs is $f_{\rm bin} \approx 1$.
By the same token, we expect up to $\sim150$, $10^3$, $10^4$ and $10^5$ binaries in LSST with times to merger of $t_m=15, 10^2, 10^3$ and $10^4$ yr, respectively.

In order to study different types of signals, we construct five sets of fiducial sources, each corresponding to a different choice of $t_m$ in $[15, 50, 10^2, 10^3, 10^4]\, {\rm yr}$.
For each value of $t_m$, we draw 100 sources from the $(M_{\rm bin},z)$ distribution in Fig.~\ref{fig:n_lsst}.
The resulting sets of binaries represent a realistic expectation of LSST sources conditioned on each $t_m$.
For simplicity, we assign all binaries a mass ratio of $q = m_2/m_1 = 0.1$, a commonly expected value for compact SMBHBs \citep{Li2020,Li2022}.

\subsection{Modeling lightcurves} \label{subsec:method-lc}

\begin{figure}
    \centering
    \includegraphics[width=\columnwidth]{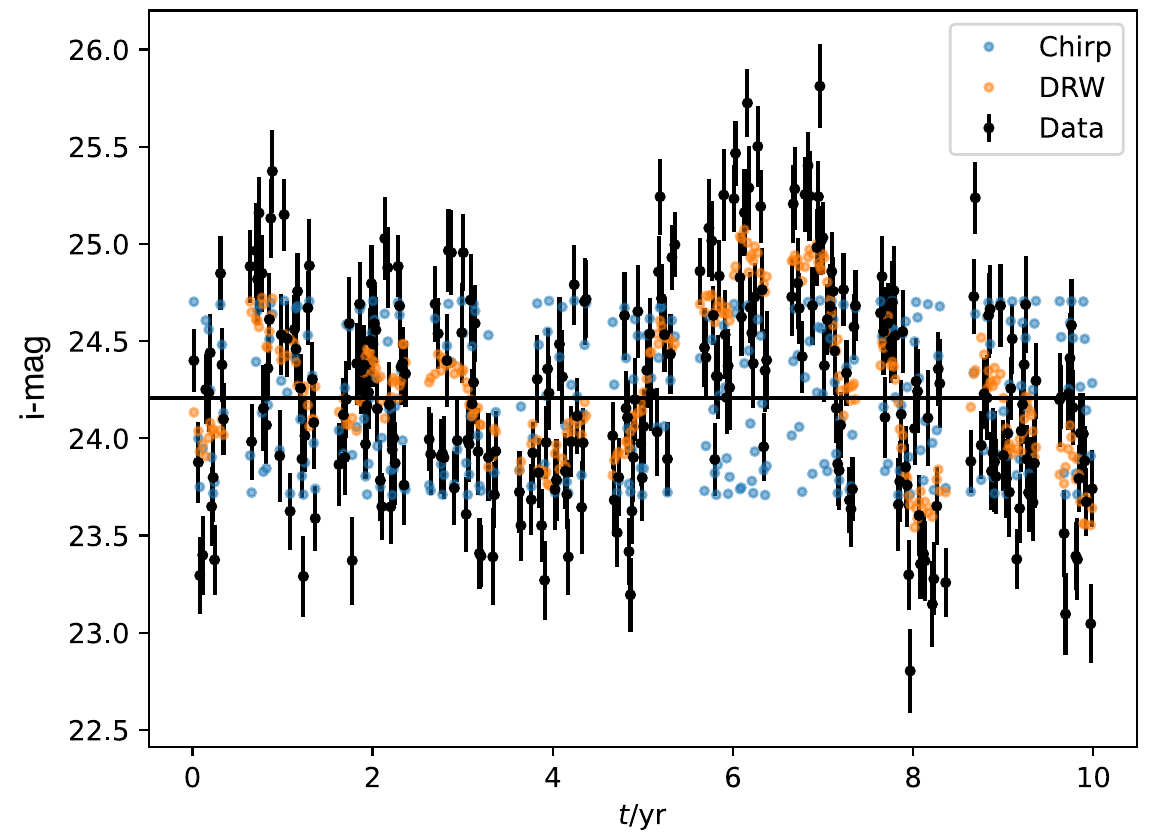}
    \caption{Example mock LSST lightcurve for a SMBHB with $\Mtot=10^{7.8} \Msun$ and $z=2.9$. The lightcurve (black points) has a mean magnitude $m_i=24.2$ per Eq.~\eqref{eq:mi} (line), modulated by the binary through relativistic DB with a chirping sinusoid, Eq.~\eqref{eq:chirp} with $A = 0.5$ mag (blue points); additionally, the quasar varies stochastically following a DRW, Eq.~\eqref{eq:drw} with $\sigma=0.5$ mag and $\tau=136$ d (orange points); finally, the photometric errors have amplitudes of $\sim 0.2$ mag (errorbars). The cadence is 6 d, with seasonal gaps that skip 1/3 of observations each year.}
    \label{fig:lc}
\end{figure} 

We next simulate $T= 10\, {\rm yr}$ of LSST data for each of the selected binaries.
Each lightcurve is centered around a mean luminosity $\mi$ given by Eq.~\eqref{eq:mi}, with modulations composed of (i) a DB chirp signal $y_{\rm chirp}$, (ii) a DRW signal $y_{\rm drw}$, and (iii) Gaussian photometric errors $y_n$, such that the overall data $y(t)$ are given by
\begin{equation} \label{eq:data}
    y(t) = \mi + y_{\rm chirp}(t) + y_{\rm drw}(t) + y_n(t) \, .
\end{equation}
Note that the first two terms on the right hand side are deterministic, and the last two are stochastic. We describe each component below, and present an example mock lightcurve in Figure~\ref{fig:lc}, for which we can measure periodicity and chirp.

A distinct periodicity in quasar lightcurves containing compact SMBHBs is expected from hydrodynamical modulations on orbital timescales, which can persist all the way through merger. The relativistic DB modulation is associated with the Doppler brightening of the emission from the mini-disks around each SMBH \citep{DOrazio2015,Charisi2022}. The DB effect introduces a modulation in flux with an amplitude of order $v/c$, where $v$ is the binary's orbital velocity. In this work, we adopt the DB model for binary variability in the LSST lightcurves. For simplicity, we assume that the SMBHB orbit is circular;
this leads to DB modulations with a sinusoidal profile \citep{DOrazio2015}. However, hydrodynamical effects \citep{Farris2015,Tang2018,Zrake2021,Westernacher-Schneider2022,Cocchiararo2024} and lensing \citep{DOrazio2018,Ingram2021,Davelaar2022a,Davelaar2022b,Krauth2024} can also cause periodic variability in SMBHBs, notably with different (non-sinusoidal) shapes and amplitudes. See \S~\ref{discussion:beyond_sin} for more details.

Assuming pure GW-driven inspiral during the last $\sim 10^4$ yr of merger, the modulation phasing is then given by a simple chirp, which we evaluate to leading post-Newtonian order \cite[e.g.,][]{Cutler:1994ys}.
Concretely, the observed (orbital) modulation frequency $f$ is related to the source-frame GW frequency $f_{\rm gw}$ by $f=f_{\rm gw}/2(1+z)$; for a binary with chirp mass $\Mc=[q/(1+q)^2]^{3/5} \Mtot$, $f_{\rm gw}$ itself evolves following
$ \dot{f}_{\rm gw} \propto \Mc^{5/3} f_{\rm gw}^{11/3}$.
We derive a reference value of the orbital frequency $f_0$ from the chosen time to merger through $t_m \propto [(1+z) \Mc]^{-5/3} f_0^{-8/3}$; this then determines the entire evolution $f(t)$.
Finally, we assume that the modulation amplitude ($A$) remains constant over the observation period, so that the form of the chirp is 
\begin{equation} \label{eq:chirp}
    y_{\rm chirp}(t) = A \cos\left[ \Phi\left(t; z, \Mc, t_m \right) - \phi\right] \, ,
\end{equation}
where the \emph{chirp phase} $\Phi$ is given by
\begin{equation}
    \Phi(t; \Mc, t_m) = - \left( \frac{t_m - t}{5 t_\Mc} \right)^{5/8},
\end{equation}
where $t_\Mc \equiv (1+z) G \Mc /c^3$ and $\phi$ is the phase at $t = t_m$.
Note that we define $t_m$, $f$, and $\dot{f}$ in the observer frame, but $\Mc$ and $f_{\rm gw}$ in the source frame.

In real signals, we expect the amplitude to increase as the binary approaches merger \citep{DOrazio2015,Haiman2017}.
Although we could build this effect into our model, we neglect it here for simplicity; an increasing amplitude should facilitate detection.

The blue points in Fig.~\ref{fig:lc} show an example of this DB-induced signal for one of our binaries in the $t_m = 50\, {\rm yr}$ set.
The initial observed frequency is $f = 3.7\times 10^{-7}$ Hz ($P = 1/f \approx 30$ d), which increases to $f \approx 4.2\times 10^{-7}$ Hz ($P=27.6$ d) over 10 years; this corresponds to a ``chirping rate'' $\gamma \equiv \dot{f} T/ f_0 \approx 7.5\%$.
For our other systems, $P$ ranges from ${\sim} 1.7$ to $300$ days, with most of them falling between ${\sim}3$ weeks to ${\sim} 3$ months.
The $\gamma$ rates are 25\%, 7.5\% 3.75\%, 0.375\% and 0.0375\% over 10 years for $t_m=$ 15, 50, 100, $10^3$ and $10^4$ years.
These binaries span the majority of the frequency domain of GW-driven SMBHBs that can be identified in LSST. 

In addition to the signal induced by orbital modulations, we expect the quasar to exhibit intrinsic stochastic variability.
A commonly used model to describe this is a DRW \citep{Kelly2009,MacLeod2010,Ivezic2013}.
Other quasar variability models have been investigated in recent years, including a ``damped harmonic oscillator" \citep[][]{Yu2022}.
We follow \citetalias{Xin2024} to implement the DRW model. The characteristic amplitude SF$_{\infty}$ and timescale $\tau$ of the DRW are empirically related to the BH mass, redshift, absolute $i$-band magnitude and rest-frame wavelength as in Eq.~(7) in \citet{MacLeod2010} or Eq.~(11) in \citetalias{Xin2024}.
We model $y_{\rm drw}(t)$ as drawn from a Gaussian process with an exponential kernel given by
\begin{equation} \label{eq:drw}
    K(t_1, t_2) = \sigma^2 \exp\left(-\frac{|t_1-t_2|}{\tau}\right) ,
\end{equation}
where ${\sigma}^2={{\rm SF}_{\infty}}^2/2$ (orange points in Fig.~\ref{fig:lc}).

Finally, we assume that the LSST observations are subject to some uncorrelated Gaussian measurement errors $y_n(t) \sim \mathcal{N}[0, \sigma_n^2(t)]$, where the variance of the Gaussian $\sigma_n^2$ is allowed to vary for each timestamp $t$.
For each datapoint, we set $\sigma_n$ randomly by drawing $\log \sigma_n \sim \mathcal{N}(\log(0.2), 0.01)$, consistent with LSST specifications \citep{LSSTScienceCollaboration2009}.
We assume that LSST's $i$-band cadence allows one visit every 6 days; to account for irregularity in the observations, we mimic a nonuniform cadence by jittering observation times by $\delta t \sim \mathcal{N}(0, (3{~\rm hr})^2)$. 
We also simulate expected ``seasonal gaps" due to the position of the Sun, which results in $\sim$1/3 of the data being blocked each year.

\subsection{Bayesian inference} \label{subsec:method-bayes}

We apply Bayesian inference to simultaneously measure the parameters of both the chirp and DRW from simulated data as in Fig.~\ref{fig:lc}.
Our likelihood is defined by the data-generating model of Eq.~\eqref{eq:data}, but with some key reparameterizations to facilitate sampling.
We parameterize the post-Newtonian chirp in terms of the orbital frequency $f_0$ and frequency derivative $\dot{f}_0$ defined at the \emph{median} LSST observation time, which are equivalent to the binary's $\Mtot$ and $t_m$.
We set priors on $f_0$ and $\dot{f}_0$ based on the expected uncertainty in the observed binary mass, which reflect the range of Eddington ratios for luminous quasars: $\Delta \log (\Mc)\approx 1$ \citep{Kollmeier+2006,Hopkins2009,Kochanek2012}, which implies
\begin{equation}
     \frac{\Delta f_0}{f_0} = - \frac{5}{11} \ \frac{\Delta \Mc}{\Mc} = 0.39\, ,
\end{equation}
    \begin{equation}
    \frac{\Delta \dot{f}_0}{\dot{f}_0} = \frac{5}{3} \ \frac{\Delta \Mc}{\Mc} = 1.43\, .
\end{equation}
It follows that $\Delta f$ is bound between min($f$) = 0.61$f_t$ and max($f$) = 1.39$f_t$, and $\Delta \dot{f}$ is bound between min($\dot{f}$) = max($0,-0.43\dot{f}_t$) and max($\dot{f}$) = 2.43$\dot{f}_t$, where $f_t$ and $\dot{f}_t$ are the true values of $f_0$ and $\dot f_0$ respectively.
In general, binaries with higher orbital frequency will have wider priors, making the computation more costly. 
Besides $f_0$ and $\dot f_0$, we infer the chirp phase $\phi$ and amplitude $A$; the model reduces to a pure DRW for $A\to 0$.
We parameterize the DRW itself by $\sigma$ and $\tau$, which are also measured from the data.

\begin{figure*}
    \centering
    \includegraphics[width=2\columnwidth]{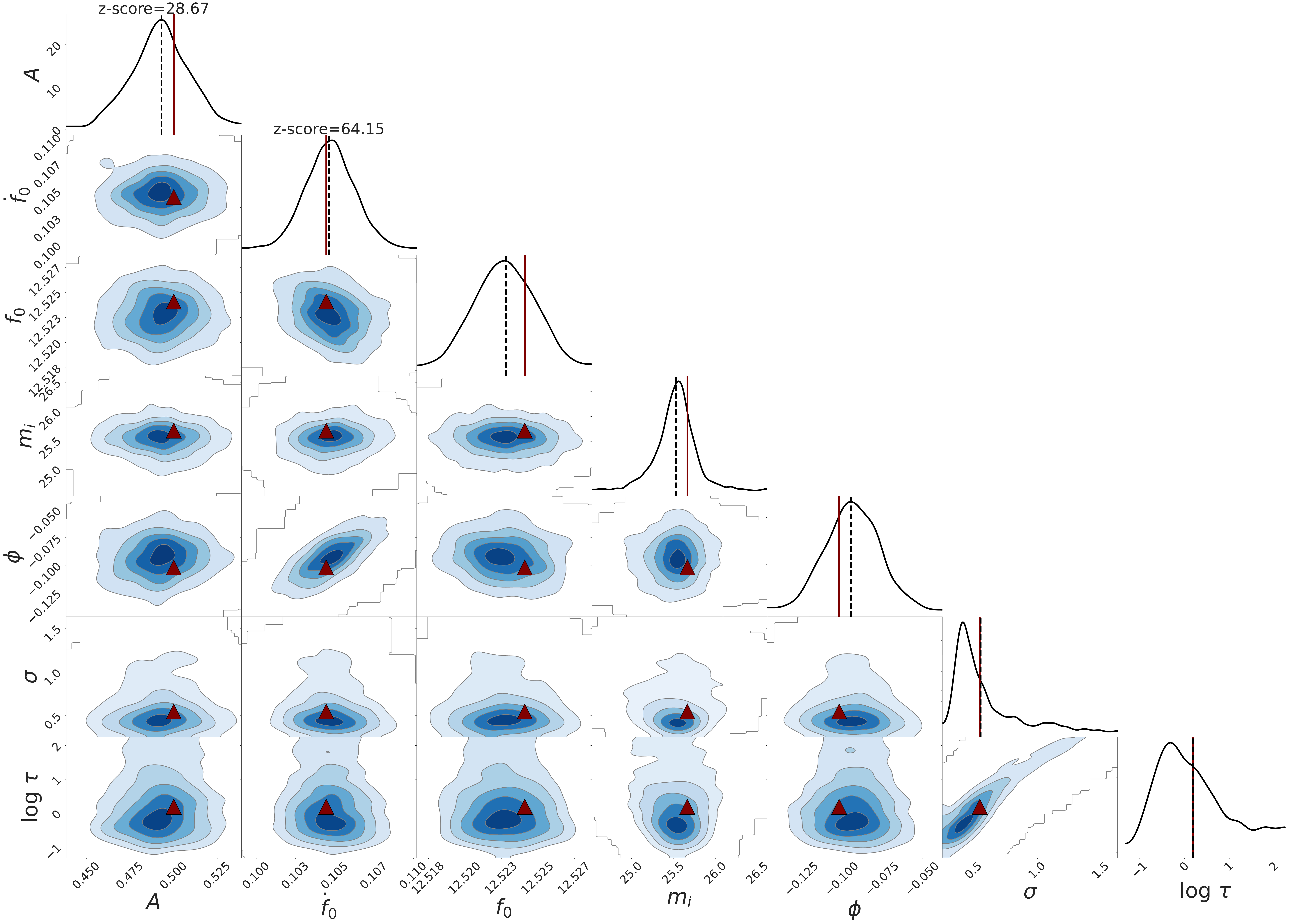} 
    \caption{Posterior for all seven parameters for the binary lightcurve in Fig.~\ref{fig:lc}. Each contour encloses 10\%, 30\%, 50\%, 70\% and 90\% of the probability respectively. We also show the truth (red triangles in the contours and red lines in the marginal panels) and mean of the posterior (black lines in the marginal) for each parameter. The $z$-score values are shown for $A$ and $\dot f_0$. 
     }
    \label{fig:corner_all_params}
\end{figure*}

The chirp likelihood is challenging to sample due to degeneracies that lead to multiple false maxima;
without dedicated treatment, this can make the sampling prohibitively inefficient.
We tackle this through a custom strategy, which we dub the ``frequency comb" method.
This involves the creation of auxiliary $f_0$ and $\dot{f}_0$ grids that are used, together with analytic marginalization over $A$ and $\phi$, in a hybrid Gibbs and Metropolis-Hastings sampling scheme.
We describe this, as well as our (generally broad) priors, in Appendix~\ref{appendix:algorithm}.

The result of the inference is a set of posterior samples on $\{f_0, \dot f_0, A, \phi, \sigma, \tau, \mi\}$. This can answer whether (i) binary modulations are detectable ($A>0$) and, if so, (ii) whether the chirp is also detectable ($\dot f_0 > 0$); we can also evaluate whether the binary parameters are recovered without bias and whether there are significant degeneracies among the binary parameters or between the noise and signal.

\section{Results} \label{sec:results}

\begin{figure*}
    \centering
    \hspace{-10mm}
    \includegraphics[width=1.1\columnwidth]{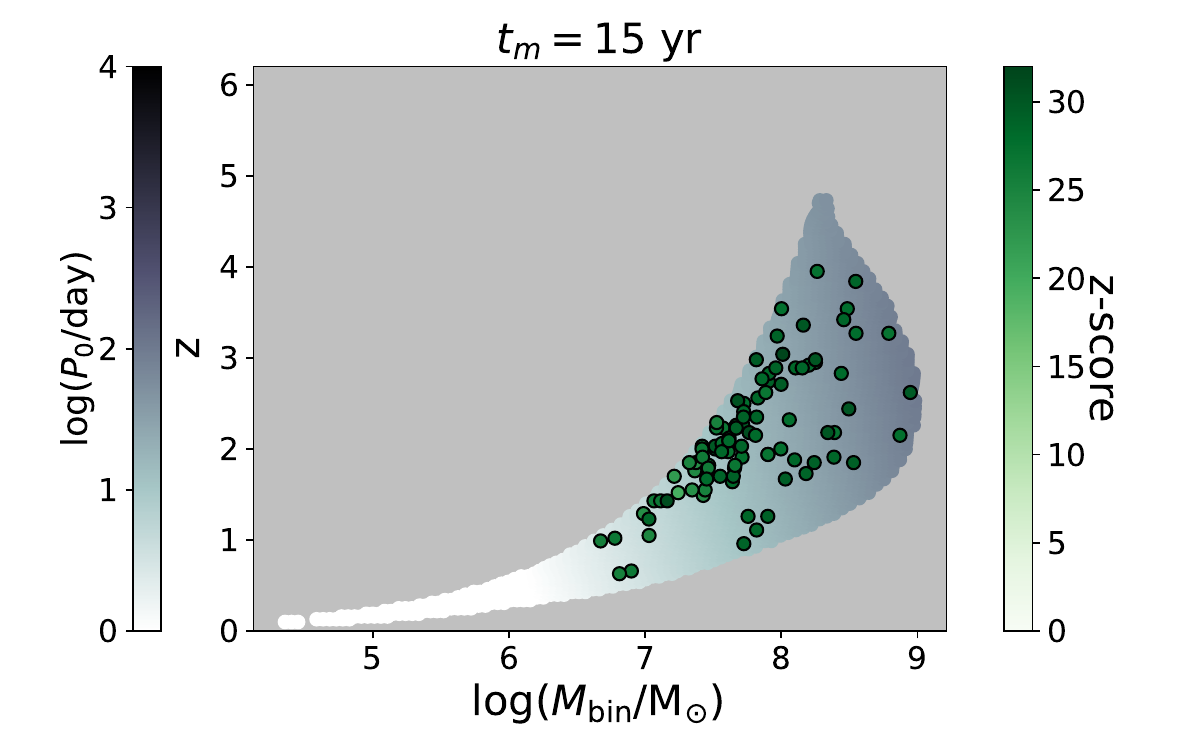} \hspace{-8mm}
    \includegraphics[width=1.1\columnwidth]{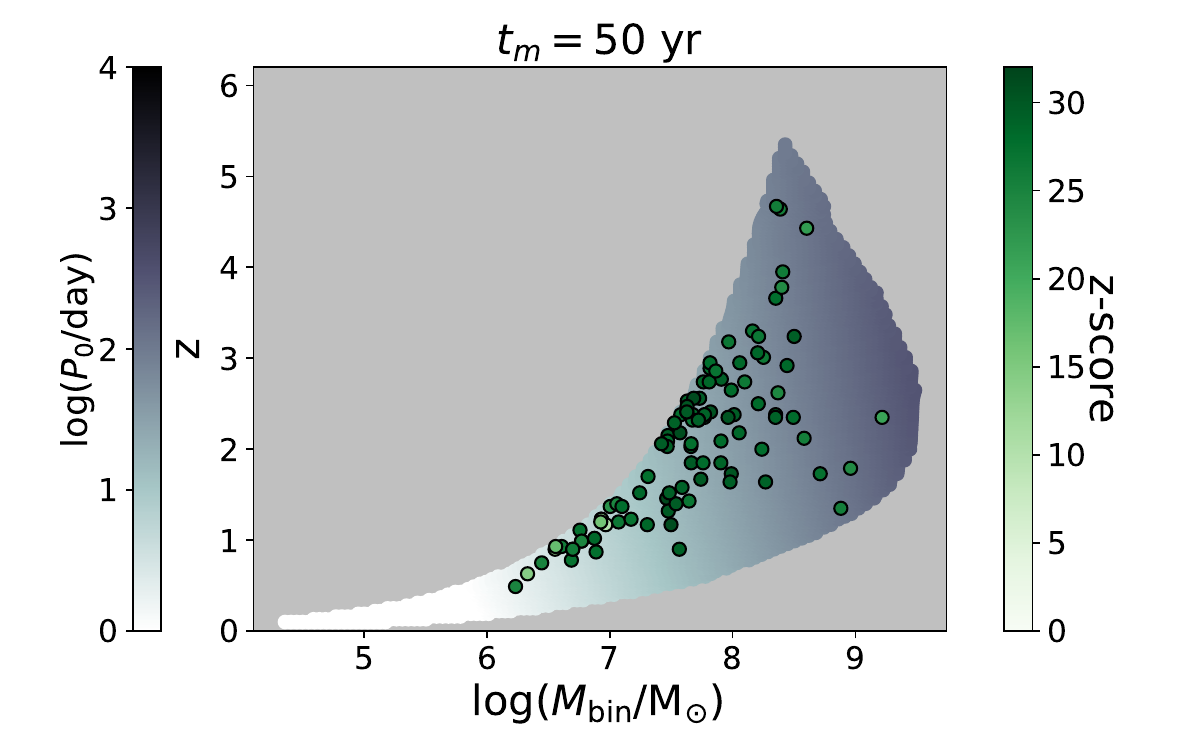} \\
    \hspace{-10mm}
    \includegraphics[width=1.1\columnwidth]{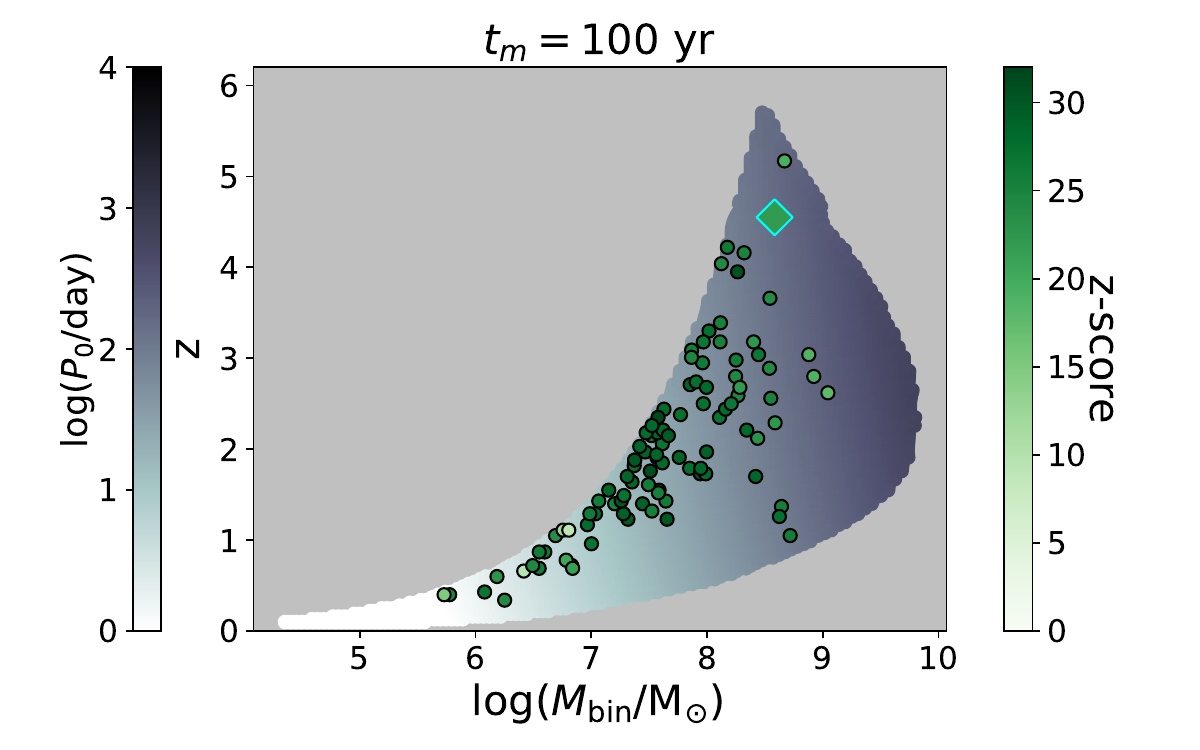} \hspace{-8mm}
    \includegraphics[width=1.1\columnwidth]{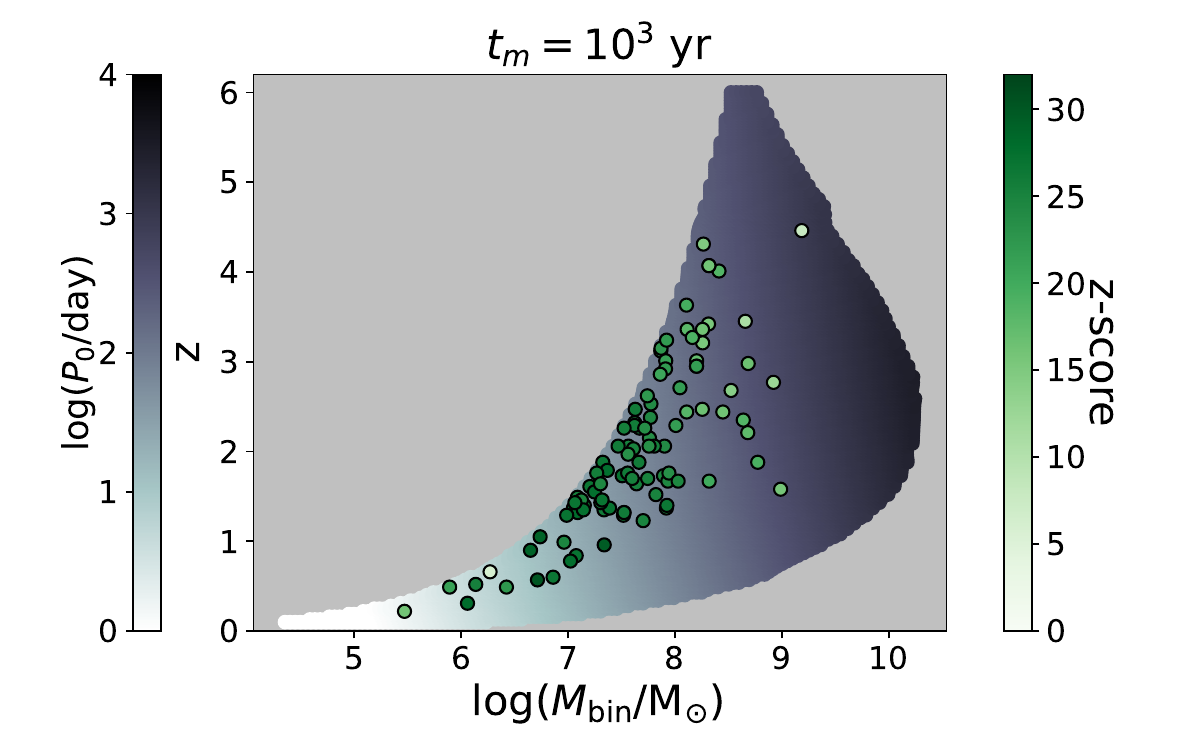} \\
    \includegraphics[width=1.1\columnwidth]{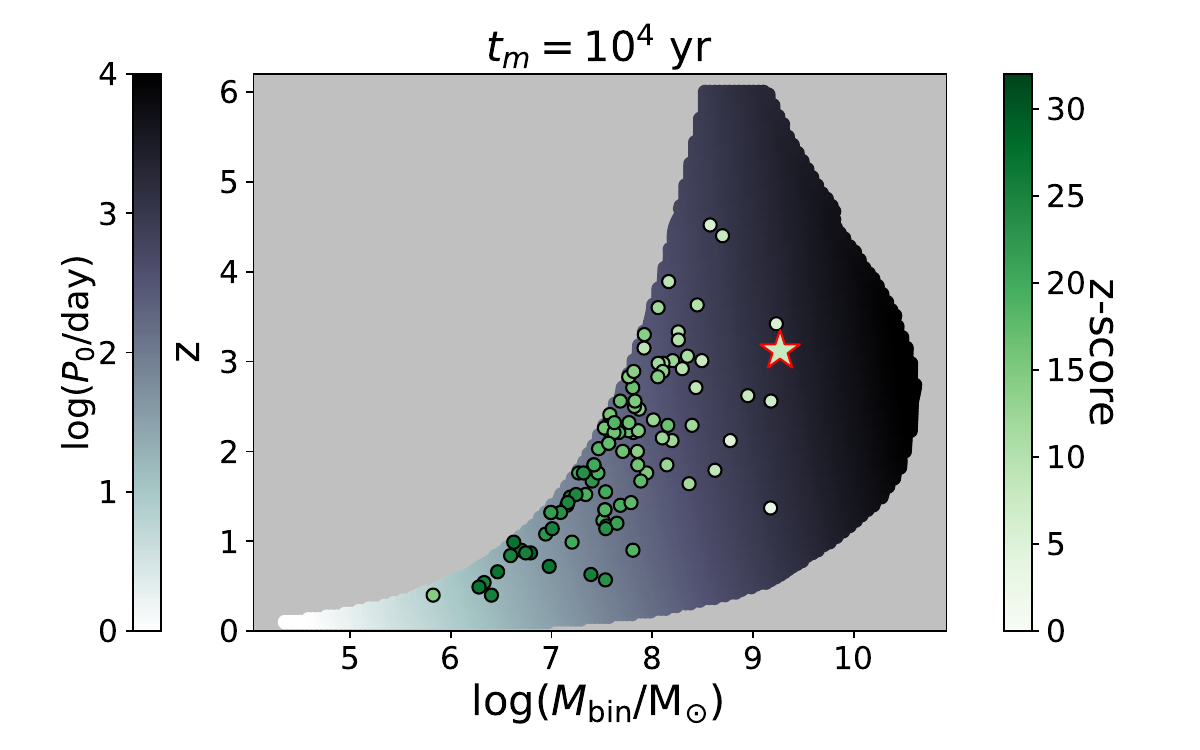}
    \caption{Summary of the inference on the 100 lightcurves simulated for  $t_m=$15, 50, 100, $10^3$ and $10^4$ yr (panels). We show posterior $z$-scores, $\mathrm{mean}(A)/\mathrm{std}(A)$, for the chirp amplitude $A$ (right colorbar), as a function of binary mass (abscissa) and redshift (ordinate), as well as the corresponding observed period $P_0 = 1/f_0$ (left colorbar). We fix the true amplitude to $A_t =$ 0.5 mag. The $z$-score is a measure of how confidently the binary signal was detected, e.g., $z\text{-score}= 5$ implies a $5\sigma$ detection.
    The star highlights the widest binary in this set that we can detect with $>5\sigma$ confidence, while the diamond highlights the widest binary for which we can detect the chirp $\dot f_0 > 0$ with $>5\sigma$ confidence.
    We present a similar plot for the $z$-score of $\dot f_0$ in Appendix.~\ref{appendix:snr_fdot}.
    } 
    \label{fig:achirp_p0}
\end{figure*}

We begin by applying our Bayesian analysis to the example lightcurve in Fig.~\ref{fig:lc}.
Figure~\ref{fig:corner_all_params} shows the resulting posterior density, compared to the true injected values.
It is apparent that the chirp is confidently detected, with both $P(A>0)$ and $P(\dot f_0 >0) > 99.98\%$, the greatest certainty achievable given our finite number of samples.
In terms of the $z$-score, i.e., the ratio of the posterior mean to its standard deviation, we determine $A>0$ at $29\sigma$ confidence and $\dot f_0 > 0$ at $64\sigma$ confidence---an unambiguous detection of the chirp.
The signal is well-recovered and the posterior shows no correlations with the DRW sector.

Assuming a linear chirp ($\dot{f} = \mathrm{const}$), the contribution of the
chirp term to the accumulated phase over an observation time $T$ is $\Delta \Phi
= \dot{f} T^2 / 2$.  Basic scaling arguments suggest that the overall phase
accuracy we can obtain in our measurement is $\sigma_{\Phi} \simeq \pi / \rho$,
where $\rho$ is the signal-to-noise ratio; in this example, $\rho = \sigma_A / A
\simeq 30$.  Ignoring correlations between $\dot{f}$ and other variables ($f$,
$A$, etc), appropriate for very loud signals, this suggests a back-of-envelope
estimate of the uncertainty on $\dot{f}$ of $\sigma_{\dot{f}} \simeq 2 \pi /
\rho / T^2 \simeq 2.3 \times 10^{-8} \, \mathrm{d}^{-2}$ in this example.  For
an average $\dot{f} \simeq 1.2 \times 10^{-6} \, \mathrm{d}^{-2}$, this
corresponds to an estimated $z$-score of $z_{\dot{f}} \simeq 52$, in close
agreement with the actual measured value.

\subsection{Effect of merger time} \label{subsec:low-high-chirp}

We next repeat the analysis on the 100 lightcurves simulated for each choice of $t_m$, still with true amplitude $A_t=0.5\, \mathrm{mag}$.
The most compact binaries considered here ($t_m=15$ yr) have periods ranging from days to weeks, making them potential future sources for LISA; the widest binaries  ($t_m=10^4$ yr) have periods of years.

Figure~\ref{fig:achirp_p0} shows that in all cases we detect the binary signal with ${>}5\sigma$ (amplitude $z$-score ${\gtrsim} 5$), and typically $10{-}30\sigma$, confidence.
The widest binary in this set whose signal we can detect (at $8\sigma$) has $P_0\approx 5$ yr ($f_0\approx 6.2$ nHz, star in bottom panel of Fig.~\ref{fig:achirp_p0}). Its chirp rate is $\gamma = 3.75\%$, corresponding to an extremely slow $\dot{f}=2.4 \times 10^{-3}$ nHz/yr; we can detect this PTA-like source with only $\sim$2 visible cycles.
As $t_m$ increases, binaries with a given mass become wider and $P_0$ increases; this leads to a reduction in the number of observed cycles and a corresponding degradation in detection confidence
as $t_m$ increases from 15 to $10^4$ yr (especially for $t_m = 10^3$ and $10^4$ yr).
On the other hand, ultrashort-period  binaries with $P_0\lesssim 10$ days can become difficult to identify due to aliasing, given LSST's ${\sim}6$-day cadence, often degrading the $z$-score despite the large number of visible cycles (evident from the `jitter' in $z$-score for such short-period sources in, e.g. the $t_m=100$ yr panel, at $z\approx 0-1$ and $\log(M_{\rm bin}/{\rm M_{\odot}})\sim 6-7$).  We expect that LSST's uneven sampling strategy (e.g. a pair of points in a single night) will ameliorate this issue.
However, for $P_0\gtrsim 30$ d, finite sampling is not an issue and the $z$-scores increase with smaller $P_0$. 

Besides detecting $A>0$, we also accurately recover the orbital frequency $f_0$ and confidently detect a non-negligible frequency evolution $\dot f_0 > 0$ for all sources in Fig.~\ref{fig:achirp_p0} (see App.~\ref{appendix:snr_fdot}).
The slowest chirp we can detect with high confidence (${>}5\sigma$) has $P_0\approx 149$ d ($f_0\approx 0.78$ nHz, diamond in middle-left panel of Fig.~\ref{fig:achirp_p0}). The $z$-score $=6$ in $\dot f_0$ and $22$ in $A$. 
Its chirp rate is $\gamma = 3.75\%$, corresponding to a slow chirp of $\dot{f}=2.9$ nHz/yr.

\subsection{Effect of binary amplitude} \label{subsec:det_chirp}

We have so far simulated $A_t = 0.5$ mag.
Both the DB model \citep{DOrazio2015} and hydrodynamical simulations \citep[e.g.,][]{Zrake2021, Westernacher-Schneider2022} can induce a range of variability amplitudes $0.1 \lesssim A \lesssim 1$ mag.
As $A$ decreases relative to the DRW, our ability to recover the chirp should degrade. Here we examine different choices of true binary amplitude, in the $0.05\, \mathrm{mag} \leq A_t \leq 0.5 \, \mathrm{mag}$ range, and study their detectability.

\begin{figure}
    \centering
    \includegraphics[width=\columnwidth]{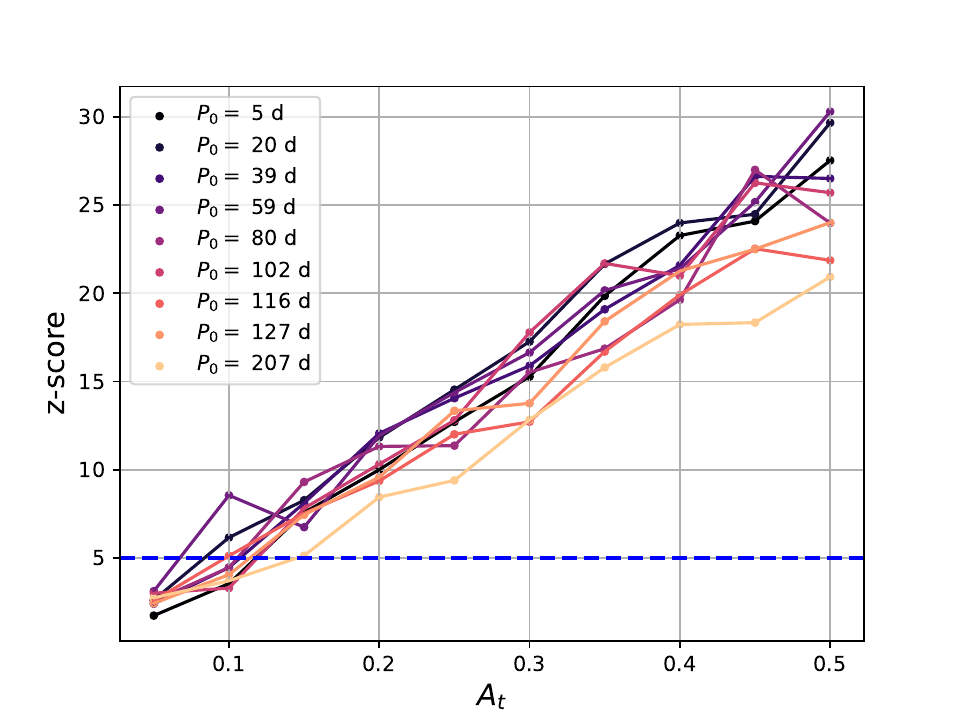}
    \caption{The detectability of binary chirp amplitude ($A$) as a function of the true injected amplitude $A_t$, for nine binaries in our sample. The $z$-score increases as $A_t$ increases. It has a weak dependence on orbital periods (quoted in the legend). The chirp rate is fixed at 7.5\% in all curves ($t_m=50$ yr). The detection is generally significant; $z$-score$ > 5$ (see blue horizontal line) when $A_t \gtrsim 0.1$.
    }
    \label{fig:zscore-Achirp}
\end{figure}

In Figure~\ref{fig:zscore-Achirp}, we show the amplitude $z$-score versus $A_t$ for binaries with $t_m = 50$ yr and varying orbital periods, $5 \, {\rm d} \lesssim P_0 \lesssim 200\, {\rm d}$.
As expected, detectability increases with the binary amplitude, reaching $z$-scores $\gtrsim5$ for $A_t>0.1$.
The $z$-score also has a weak dependence on $P_0$, and is slightly higher for binaries with shorter $P_0$, consistent with the trends in Fig.~\ref{fig:achirp_p0}. 
\begin{figure*}
    \centering
    \includegraphics[width=2\columnwidth]{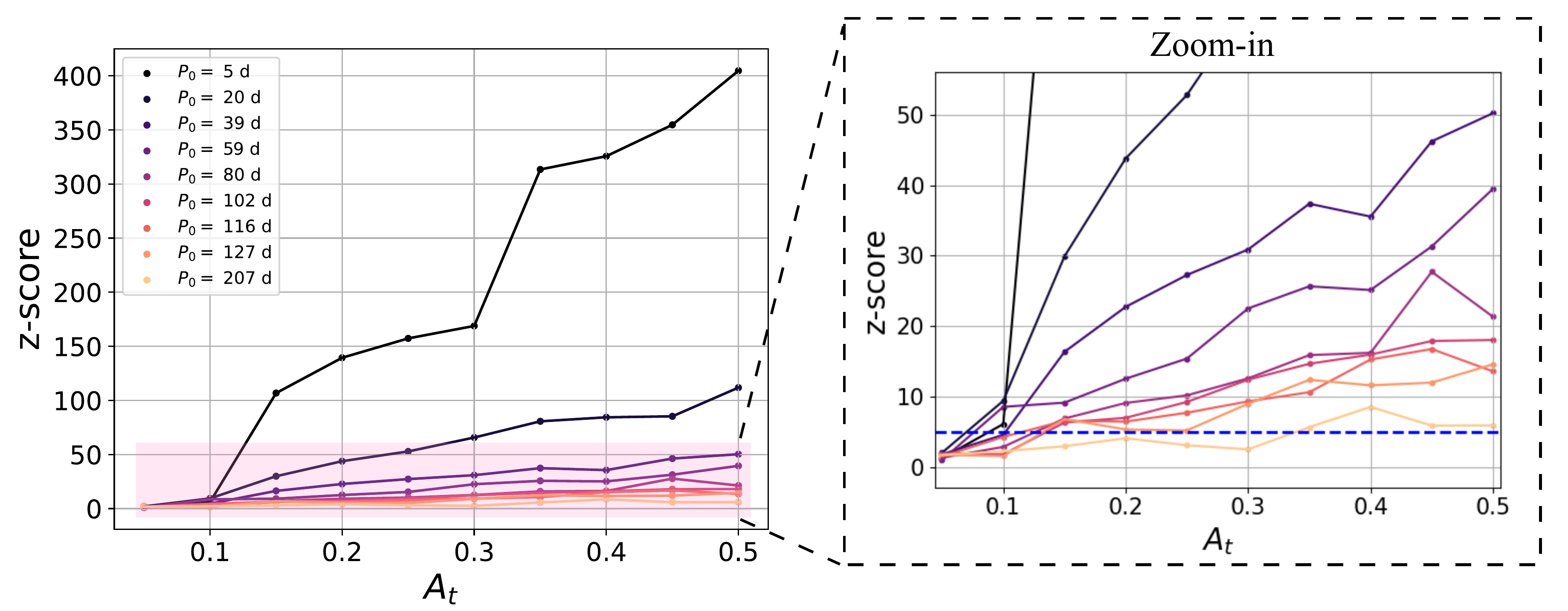}
    \caption{The significance at which the frequency chirp $\dot{f}_0$ can be measured as a function of variability amplitude $A_t$ (abscissa) and period $P_0$ (color). The y-axis shows the posterior $z$-score for $\dot{f}_0$, i.e., $\mathrm{mean}(\dot f_0)/\mathrm{std}(\dot f_0)$, calculated for the same set of $A_t$ and $P_0$ as in Fig.~\ref{fig:zscore-Achirp}, i.e., with chirp rate $\gamma = 7.5\%$ ($t_m=50$ yr). We achieve overall large $z$-scores, meaning that $\dot{f}_0$ can be inferred at high confidence, especially when $A_t\gtrsim 0.2$. The $z$-score is higher when $P_0$ is shorter, because more cycles are observed and the period changes more rapidly. }
    \label{fig:zscore-wdot}
\end{figure*}

To further assess the detectability of the chirp, Fig.~\ref{fig:zscore-wdot} reports $z$-scores in $\dot{f}$, i.e., posterior $\mathrm{mean}(\dot f_0)/\mathrm{std}(\dot f_0)$, for the same simulations as Fig.~\ref{fig:zscore-Achirp}.
We find that we can measure the frequency growth with high confidence (${\geq}5\sigma$) for all but the weakest signals in this set.
The weakest $\dot f_0$ we can detect corresponds to a binary with $P_0\approx 200$ d ($f_0=47.9$ nHz) and $A_t \geq 0.3$ mag (red curve in zoom-in);
this binary has $\gamma = 7.5\%$, and $\dot{f}=0.43$ nHz/yr.
We can measure $\dot f_0 >0$ more confidently with decreasing $P_0$, achieving ${\geq}5\sigma$ confidence for all amplitudes of interest ($A_t \geq 0.1$ mag) starting with $P_0 < 116$ d.
This is expected from the number of visible cycles, per the $t_m=50$ yr trend in \S\ref{subsec:low-high-chirp}.

\subsection{Inferring binary properties} \label{subsec:mc-tc}

\begin{figure*}
    \centering
    \includegraphics[width=0.67\columnwidth]{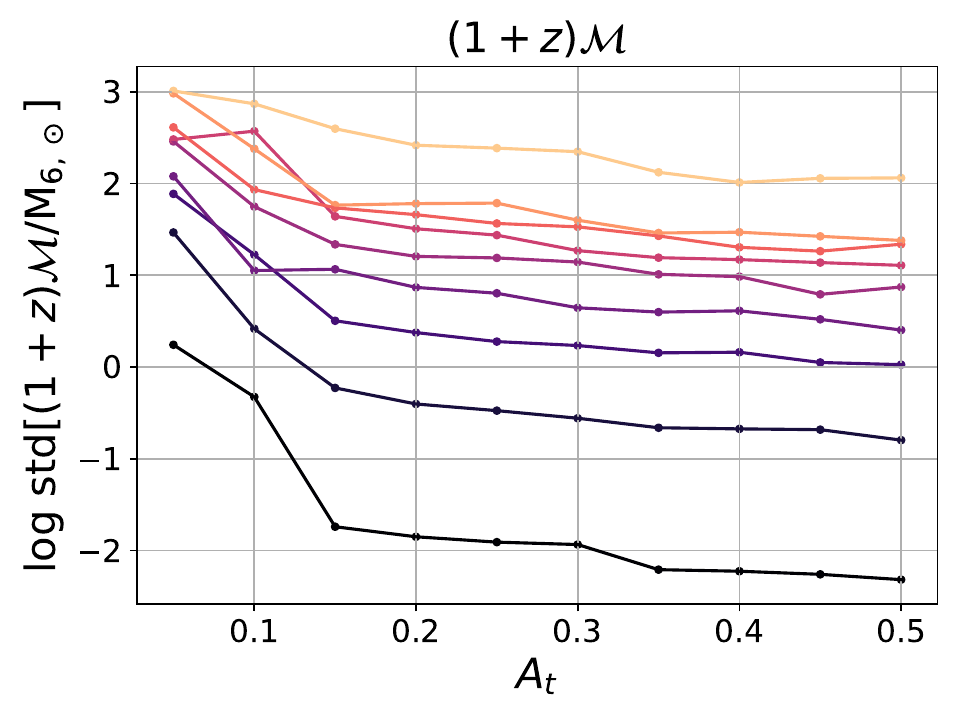} 
    \includegraphics[width=0.67\columnwidth]{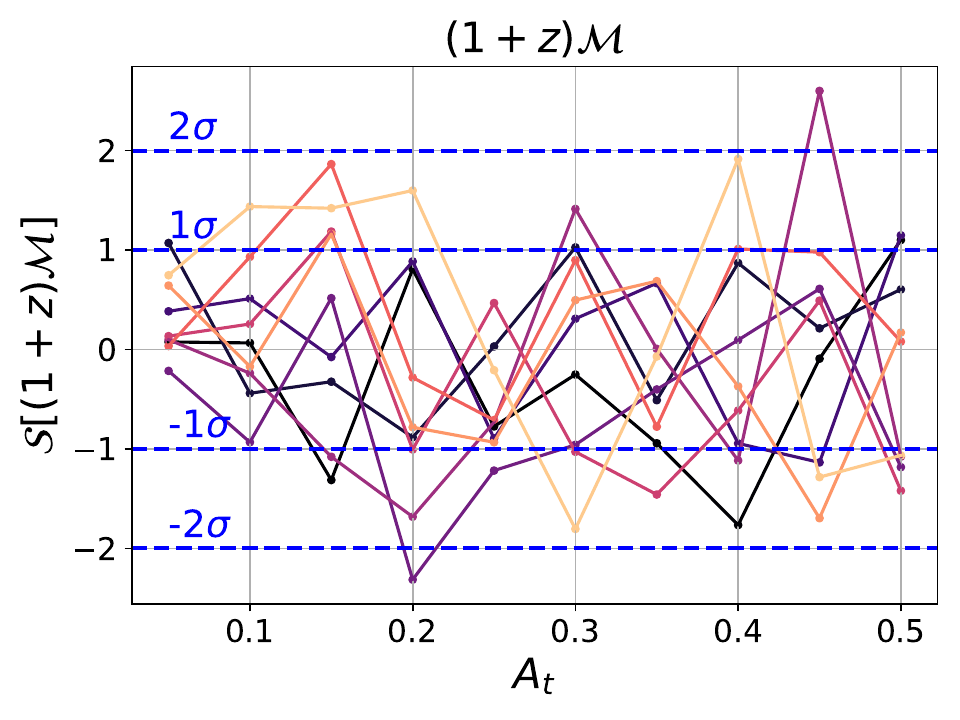} 
    \includegraphics[width=0.67\columnwidth]{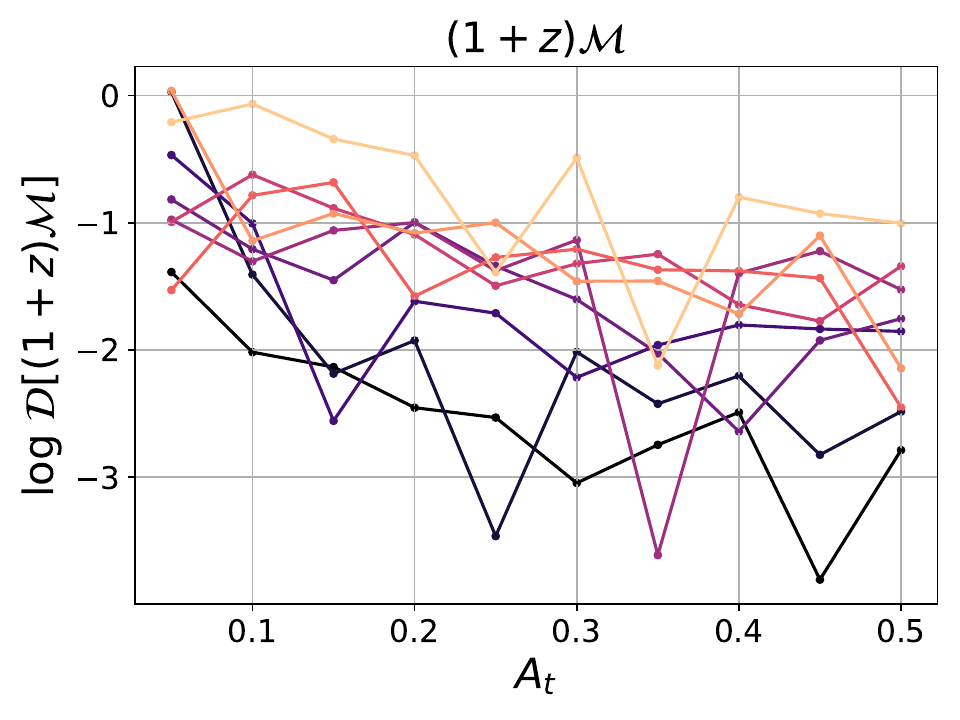} \\
    \includegraphics[width=0.67\columnwidth]{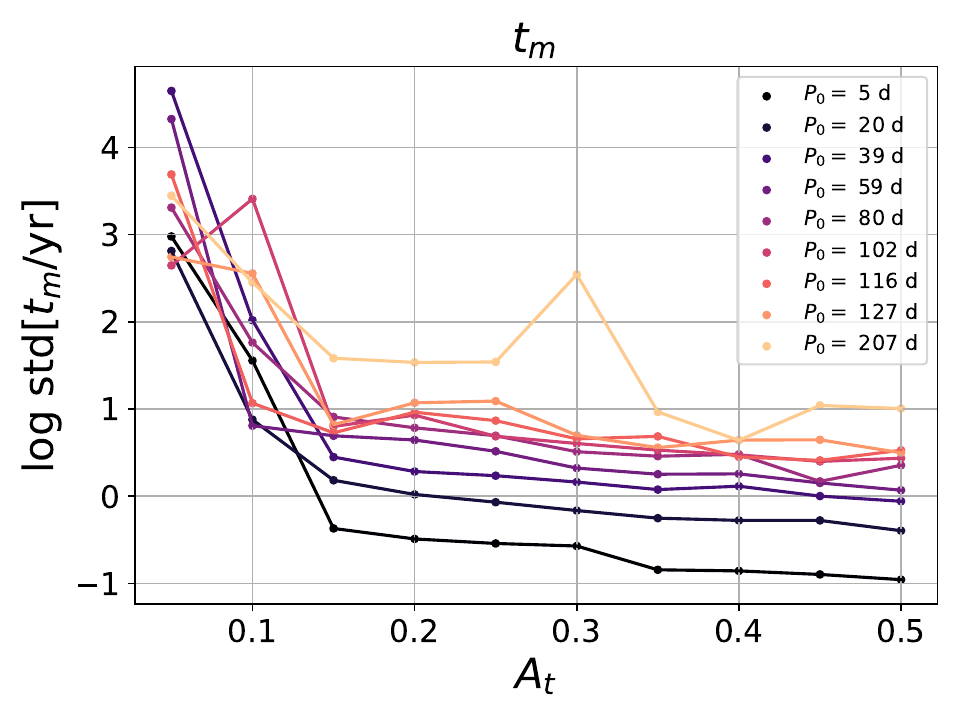} 
    \includegraphics[width=0.67\columnwidth]{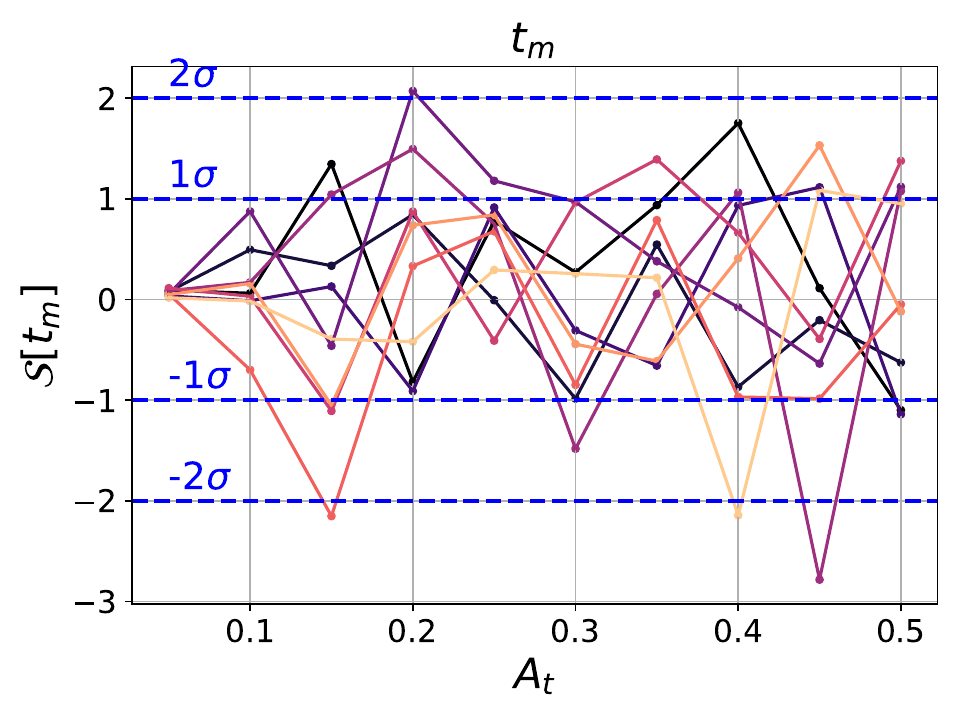} 
    \includegraphics[width=0.67\columnwidth]{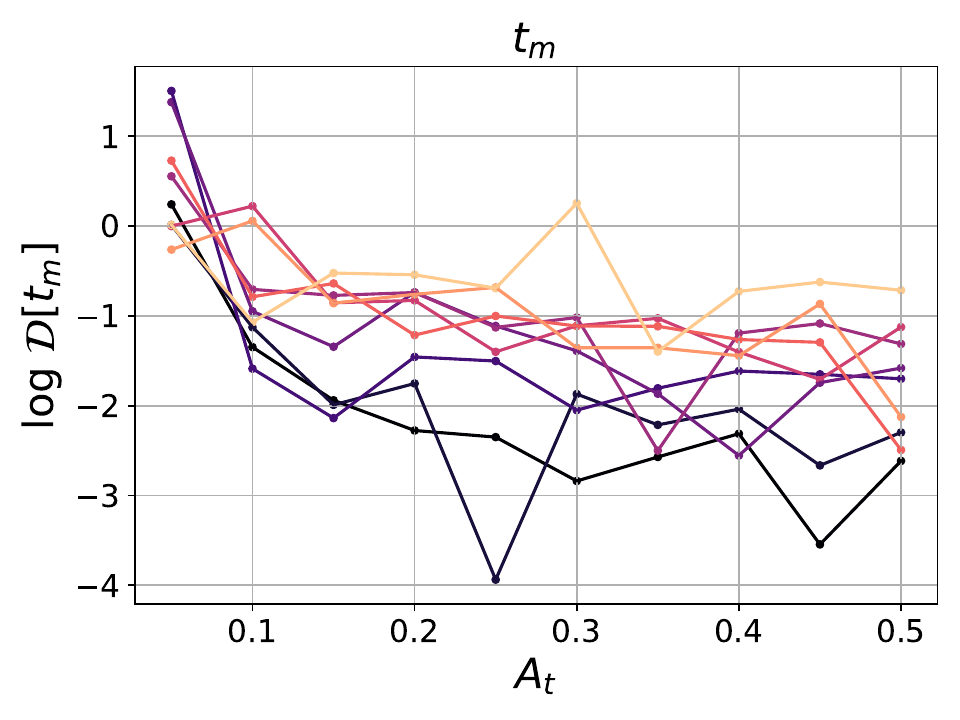} 
    \caption{Precision and accuracy with which we can measure the observed chirp mass
    $\McDet$ (top left panel) and time-to-merger $t_m$ (bottom left panel), as a function of the true binary amplitude
    $A_t$ (abscissa) and period $P_0$ (color).
    The left column shows precision through log-${\rm std}(x)$ in units of ${\rm M_{6,\odot}}=10^6{\rm M_{\odot}}$ for [$(1+z)\mathcal{M}$] and yr for [$t_m$].
    The middle column shows accuracy through the
    posterior mean distance in units of standard deviation, as in Eq.~\eqref{eq:st_err}; dashed lines mark $1\sigma$ and $2\sigma$ deviations. The right column shows the fractional error as defined in Eq.~\eqref{eq:d_err}.
    } 
    \label{fig:zscore-mc-tc}
\end{figure*}

Physical properties of the SMBHBs, such as $t_m$ and the redshifted chirp mass, $\McDet$, can be derived from our inference of $f_0$ and $\dot f_0$ (\S\ref{subsec:method-lc}).
As for the signal parameters $\{A, \phi, f_0, \dot f_0\}$, the precision with which we can recover $\McDet$ and $t_m$ improves with signal-to-noise ratio, as proxied by $A_t$.
In Fig.~\ref{fig:zscore-mc-tc}, we show how the \emph{precision} in both parameters improves as functions of binary amplitude $A_t$ (left column) for the same binaries as in Fig.~\ref{fig:zscore-Achirp}.
The uncertainty in the measured $(1+z)\mathcal{M}$ can be at best ${\sim}10^4 {\rm M_{\odot}}$ (obtained for $P_0=5$ d binaries and $A_t > 0.15$).
For some binaries ($A_t > 0.3$ and $P_0 < 200$~d), we can constrain $t_m\lesssim 5$~yr.
Assuming conservatively that LISA is launched without significant delay in the mid-2030s, coinciding with the end of the 10-yr LSST observations, and that LISA operates for 5 years, these binaries will merge during LISA's observational window.
They can thus be flagged 
as excellent candidates for a targeted search with LISA prior to the start of the LISA observations.

Additionally, we show the \emph{accuracy} of our measurements by plotting the deviation of the posterior mean from the truth relative to the measurement precision (middle column), i.e.,
\begin{equation}
    \mathcal{S}[x] = \frac{\mathrm{mean}(x) - x_t}{\mathrm{std}(x)}\, , \label{eq:st_err}
\end{equation}
for $x = \McDet$ or $t_m$. 
The mean displacements from the truth are typically $<2\sigma$ irrespective of
$A_t$ and $P_0$, and they are roughly equally spread in the positive and negative directions, suggesting our measurements are reasonably unbiased. 
The exceptions occur at minimum $A_t=0.05$, where $\mathcal{S}[x]$ plunges to $\sim 0$ for both $(1+z)\mathcal{M}$ and $t_m$. This is because those binary signals are too weak to be detected: their posteriors are close to the prior, whose mean lands at zero close to the truth relative to the wide variance. Finally, in right column, we show the \emph{fractional error} (defined in Eq.~\ref{eq:d_err}) for measuring $(1+z)\mathcal{M}$ and $t_m$. 
\begin{equation}
    \mathcal{D}[x] = \frac{|\mathrm{mean}(x) - x_t|}{x_t}\,. \label{eq:d_err}
\end{equation}
For most binaries, both $(1+z)\mathcal{M}$ and $t_m$ can be measured with $\lesssim$10\% fractional errors, except the widest binaries ($P_0=200$ d) and the smallest amplitudes ($A<0.2$). 

\section{Summary and Discussion} \label{sec:discussion}

In this paper, we have developed a Bayesian inference framework to measure seven parameters associated with a GW-driven SMBHB chirp signal in LSST quasar lightcurves. We find that chirps from a GW-driven inspiral are typically detectable (both in amplitude and frequency derivative) with high confidence for realistic choices of the SMBHB parameters.
The detection significance depends on the signal's amplitude and period, which control the signal-to-noise ratio and the chirp rate.  The widest system for which we can detect the chirp signal ($A$) has orbital period of $P_0 \approx 1850$ d, and chirps extremely slowly with $\dot f_0 \approx 2.3\times 10^{-3}$ nHz/yr; the widest binary for which we can measure chirp ($\dot f_0\neq 0$) {\it well} is one with $P_0 \approx200$ d and $\dot{f}=0.43$ nHz/yr.

Detecting a chirp represents smoking-gun evidence for a compact SMBHB.
Some of the SMBHs detectable by LSST will also be potential LISA sources (day to week periods). 
Identifying LISA binaries as chirping quasars in LSST can place important empirical constraints on the expected binary merger rates before LISA, and enable multimessenger observations once LISA launches.
We expect at least $\sim$145,000 compact SMBHBs in LSST with orbital periods of $P=1850$ days or shorter, or with a chirp of $2.3\times 10^{-3}$ nHz/yr or faster.
Our results show that we should be able to measure non-zero chirp from all of these binaries, assuming typical amplitude variations of $\sim 0.1-0.5$ mag.

According to our assumption that a large fraction of luminous quasars are associated with SMBHBs, we expect that the sources making up the stochastic GW background (GWB) detected by PTAs will be in the LSST quasar catalog.  \citet{Kis-Toth2025} showed that the GWB in the nHz bands is dominated by compact $\sim 10^{9} {\rm M_{\odot}}$ binaries at $z=2-3$ (see their Figure 5), with a non-negligible contribution from wider binaries with masses as low as $\sim 10^{6} {\rm M_{\odot}}$. Our results imply that the bulk of the sources of the GWB will be possible to identify from EM observations in LSST alone.

Our method represents a significant improvement over standard techniques, like the Lomb-Scargle periodogram, which has difficulty with chirping signals.
It achieves this at an acceptable computational cost, taking typically 10 min per lightcurve (Appendix~\ref{appendix:computational-cost}).
Since our method can detect chirps across a wide range of binaries, with time-to-merger up to $10^4$ yr, existing SMBHB candidates are suitable subjects for our study. In future work, we will search for periodic quasars in current time-domain surveys, including CRTS and PTF, and revisit existing candidates. 

However, our analysis is subject to some key assumptions, including the binary variability shape and the noise model. In this section, we emphasize future steps to improve our analysis, particularly incorporating other binary models, quantifying any false alarm rates and using realistic LSST data.

\subsection{Beyond the sinusoid profile} \label{discussion:beyond_sin}

A notable SMBHB candidate, PG1302-102, has established periodicity that may be explained by DB modulations, both in its sinusoidal variability observed in the optical band \citep{Graham2015b} and in its features observed in multi-wavelength studies \citep{DOrazio2015, Xin2020}. Beyond the simple sinusoidal model motivated by DB variability for a circular orbit, the periodic variability of a SMBHB can have more complex profiles due to several effects: DB in eccentric orbits, hydrodynamical effects \citep{Farris2015,Zrake2021,Westernacher-Schneider2022,Cocchiararo2024}, self-lensing flares \citep{DOrazio2018,Ingram2021,Davelaar2022a,Davelaar2022b,Krauth2024}, or, more realistically, a combination thereof \citep{DOrazio2024}. 

Hydrodynamical simulations have shown that thermal emission from the disk can produce a saw-toothed or spikey periodicity.
Self-lensing flares occur when the binary orbit is aligned with the line-of-sight within the Einstein radius, typically producing spikey-shaped periodicity as one SMBH (in the foreground) magnifies the emission from the second (background) SMBH during its passage.
All these types of variability are subject to chirping during the GW-driven inspiral stage, making the chirp a robust observable.

We will extend our Bayesian framework to incorporate these different profiles, in particular the saw-toothed and spike pulse shapes. This is similar to the work by \cite{Davis2024} and \cite{Park2024}, who have conducted population synthesis of saw-toothed lightcurves and self-lensing flare lightcurves in LSST (however both without chirping frequency).
\cite{DOrazio2024} have developed a public tool (\texttt{binlite}) to fit lightcurves, using LSST specifications, with a combination of DB, hydrodynamical, and lensing models. In future work, we will apply our method to such complex periodic and chirping lightcurves.

In our work, for simplicity, we assume that SMBHBs are inspiraling due to GW emission alone.   It is possible that the circumbinary gas, which we are assuming produces the periodic EM emission influences the inspiral rate.  The analytic models in \citet{Haiman2009a}, as well as recent hydrodynamical simulations \citep{Dittmann2023,Garg2024,ONeill2025} imply that compact binaries, even inside the LISA band, can still be coupled to their circumbinary disks, especially if the disk viscosity is high and/or the binary is retrograde with respect to the disk.  In this case, the inspiral is generally expected to be accelerated compared to the pure GW-driven case, which means that it could be more observable than our results imply.  On the other hand, since these gaseous effects are uncertain, additional free parameters will be required in the LSST lightcurve analyses to allow for these uncertainties.   We leave investigations of these issues to a follow-up study.

\subsection{False positives from non-chirp variations} 

In this work, we follow the chirp template as predicted by the GW frequency evolution. However, other forms of chirp-like frequency evolution might arise from the lightcurves that are not caused by a binary SMBH. For example, the only reported binary candidate with an observed rate of change in its orbital period is the so-called ``tick-tock" binary \citep{Jiang2022}. However, \citet{Dotti2023} have ruled out ``tick-tock" as a chirping SMBHB in optical follow-ups; they have found that its frequency change might be explained by AGN-disk precession around a single SMBH. For this reason, we will implement alternative templates for $\dot{f}$ that do not follow the GW frequency, and we will assess whether or not our Bayesian model can distinguish them from true GW chirps. These templates be (i) an arbitrary non-zero constant $\dot{f}_0$ deviating from $\dot{f}_{\rm gw}$, or (ii) setting the higher order ($\mathcal{O}(\geq2)$) derivatives of $\dot{f}_{\rm gw}$ to be zero.
Additionally, the circumbinary disk can drive a SMBHB to mildly eccentric \citep[$e\sim 0.4-0.5$; ][]{Zrake2021,Siwek2023} or highly eccentric \citep{Tiede2024} orbits, resulting in a different GW chirp compared to circular case \citep{P.C.PETERS1963}.

\subsection{Stochastic AGN variability and false alarm rate}
We have adopted the commonly-used DRW model to describe the stochastic variability of AGN in the optical band \citep{Kelly2009, MacLeod2010}, in our study. \citet{Vaughan2016} have shown that pure DRW noise can mimic a periodic signal, given a sufficiently large sample of pure DRW lightcurves. We emphasize that even though we produce only one DRW realization per lightcurve, our posterior statistics already demonstrate the probabilities that the truth can be found inside a certain credible region (see Fig.~\ref{fig:corner_all_params}); generating more DRW realizations will give very similar results. 

However, we assume that in real life, the binary and noise variability conform to our model -- DB + DRW. Under these assumptions, we detect a wide range of binary signals (i.e. in Fig.~\ref{fig:achirp_p0}) with very high confidence; z-score's are between $\sim 8-30$, corresponding to extremely small false alarm probability of $\lesssim 10^{-16}$. 
On the other hand, both the binary model (e.g discussion in \S~\ref{discussion:beyond_sin}) and the noise model are inherently uncertain, and these uncertainties will dominate the z-score, precluding such low false alarm probabilities.  Various works have demonstrated the shortcome of DRW on fitting the high-frequency end of quasar variability \citep[e.g.,][]{Mushotzky2011,Smith2018} due to sparseness of the AGN data used in the original studies. Alternative models have been proposed including the so-called ``damped harmonic oscillator" \citep{Yu2022}, which has found different correlations between the timescale and amplitude of AGN noise and the physical parameters (BH mass, luminosity and wavelength).

Although we do not explore these other noise models here, we note that, if the observation baseline is long enough ($T\gg \tau$), and there are many chirp cycles in the observation, then the precise description of the noise model should not matter much (the chirp and noise signals become uncorrelated).
Previous studies analyze data from higher-mass quasars, with masses $M_{\rm bh}=10^{8-10} \Msun$, due to the observing limitations of the AGN lightcurves. With LSST, we expect low-mass AGNs, down to $M_{\rm bh}\sim 10^5\Msun$, to be the most common. Therefore, detections of low-mass AGNs with LSST will be essential for understanding AGN intrinsic variability.

We plan to investigate alternative noise models, including using realistic AGN lightcurves.
However, low-mass AGN observations would be needed to empirically determine realistic noise for a significant fraction of binaries considered in our study (currently unavailable).

\subsection{LSST multi-band data} 

LSST observes 6 optical bands on a rolling basis, with each band revisited every ${\sim} 2$ weeks \citep[although some can be visited more frequently,][]{LSSTScienceCollaboration2009}. In this work, we only consider the $i$-band, which can have cadence as short as 6 days. Our method performs well on chirp signals for ultra-compact binaries that have periods down to $P_0\sim1$ day, which is a few times above the Nyquist frequency. However, as we discussed in \S\ref{subsec:low-high-chirp}, some shorter-period binaries experience aliasing systematics when the orbital periods are closer to Nyquist. Combining different bands and shortening the effective cadence can potentially improve our statistics for short-period sources. This can be achieved assuming constant color correlations between each LSST band, and combining all bands can reduce the effective cadence to be as low as $\sim$3 days. 

\begin{acknowledgments}
  ZH thanks the Center for Computational Astrophysics (CCA) at the Flatiron Institute for their hospitality during a sabbatical stay, where this work was initiated.  The Flatiron Institute is a division of the Simons Foundation.    This work was supported in part by NASA's LISA Preparatory Science Program (LPS) through grant 80NSSC24K0440 and by NASA Astrophysics Theory Program (ATP) grant 80NSSC22K0822.
\end{acknowledgments}

\software{\textsc{numpyro} \citep{phan2019composable,bingham2019pyro};
\textsc{celerite2} \citep{celerite1,celerite2}};
\textsc{Python} \citep{python};
\textsc{The Software Citation Station} \citep{software-citation-station-paper, software-citation-station-zenodo}.

\appendix

\section{Bayesian Algorithm} \label{appendix:algorithm}

The probabilistic model that defines our inference is as described in \S\ref{subsec:method-bayes} with the following specifications.
As in Eq.~\eqref{eq:data}, we assume that the data are given by both deterministic and stochastic contributions.
The deterministic component is comprised of the mean $m_i$ and the DB chirp $y_{\rm chirp}(t)$, as in Eq.~\eqref{eq:chirp}.
The stochastic component is given by a DRW $y_{\rm drw}(t)$ with exponential spectrum as in Eq.~\eqref{eq:drw}, as well as uncorrelated, zero-mean Gaussian noise with variances $\sigma_{n,k}^2$ for each timestamp $t_k$ due to photometric errors.
Conditional on a choice of $\sigma_{n,k}^2$'s as well as the DRW amplitude $\sigma$ and correlation length $\tau$, the likelihood is thus a zero-mean Gaussian on the residuals,
\begin{equation}
r(t) = y(t) - m_i - y_{\rm chirp}(t; A, \phi, f_0, \dot f_0)~,
\end{equation}
with covariance matrix $C(\sigma, \tau; \sigma_n)$ given by the DRW spectrum of Eq.~\eqref{eq:drw}, with diagonals enhanced by $\sigma_{n,k}^2$ \citep{2006gpml.book.....R}.
Equivalently, for $N$ observations, the data-generation process is given by an $N$ dimensional Gaussian,
\begin{equation} \label{eq:data-generation}
    y \sim \mathcal{N}[m_i + y_{\rm chirp}(A, \phi, f_0, \dot f_0), C(\sigma, \tau; \sigma_n)]
\end{equation}
where the covariance matrix $C$ is $N\times N$ and $y$, $y_{\rm chirp}$ and $\sigma_n$ are all of length $N$.
We evaluate the DRW likelihood function, $p(y \mid m_i, A, \phi, f_0, \dot f_0, \sigma, \tau) = \mathcal{N}[y_{\rm chirp} + m_i, C(\sigma, \tau; \sigma_n)](y)$, in a numerically efficient way using \textsc{celerite2} \citep{celerite1,celerite2}.

Together with a choice of prior, the likelihood of Eq.~\eqref{eq:data-generation} defines the posterior over the seven free parameters in our model, i.e., $\{f_0, \dot f_0, A, \phi, \sigma, \tau, m_i \}$, or any reparameterization thereof. 
We use \textsc{numpyro} \citep{phan2019composable,bingham2019pyro} to draw from the 7D posterior through a hybrid Hamiltonian Monte Carlo (HMC) and Gibbs sampling strategy, together with analytic marginalization over linear parameters.
This strategy is designed to address two major challenges:
\begin{enumerate}
    \item the chirp phase, frequency and frequency derivative can be near degenerate over short periods of observation;
    \item it is inefficient to directly sample the chirp amplitude $A$ and phase $\phi$ as written in Eq.~\eqref{eq:chirp}.
\end{enumerate}
We develop a bespoke solution for the former, and apply a pre-existing linearization and factorization strategy for the latter.
We detail this next, describing our priors along the way (summarized in Table \ref{tab:priors}).

\begin{table}[tb]
\centering
\caption{Prior distributions for all explicitly sampled quantities; definitions clarify inputs to each prior.\footnote{$\mathcal{N}(x,y)$ is a normal with mean $x$ and variance $y$, $\mathcal{U}(x,y)$ is a uniform distribution between $x$ and $y$, $\mathcal{N}_{[\,0,\infty)}(x,y)$ is a normal distribution truncated to the positive half-line, and $\mathcal{U}\{j,k\}$ is a discrete uniform distribution over the integers from $j$ to $k$, inclusive.}
}
\label{tab:priors}
\begin{tabular}{llll}
\hline
\textbf{Parameter} & \textbf{Meaning} & \textbf{Prior Distribution} & \textbf{Definitions} \\
\hline
$\log\sigma$
    & Log DRW amplitude
    & $\displaystyle \mathcal{N}\bigl(\log\sigma_{y},\,4\bigr)$
    & $\sigma_{y} = \mathrm{std}(y)$ lightcurve standard deviation\\[6pt]
$\log\tau$
    & Log DRW timescale
    & $\displaystyle \mathcal{U}\bigl(\Delta t,\,T\bigr)$
    & $\Delta t = \text{median cadence},\; T = \text{total span}$ \\[6pt]
$w_{0}=2\pi f_{0}$
    & Anchor for $\omega_{0}$ grid
    & $\displaystyle \mathcal{N}_{[\,0,\infty)}\bigl(\omega_{\min},\,\Delta\omega^{2}\bigr)$
    & $\omega_{\min}=\text{min. frequency},\; \Delta\omega = \text{grid spacing}$ \\[6pt]
$\dot{w}_{0}=2\pi \dot{f}_{0}$
    & Anchor for $\dot{\omega}_{0}$ grid
    & $\displaystyle \mathcal{N}_{[\,0,\infty)}\bigl(\dot{\omega}_{\min},\,\Delta\dot{\omega}^{2}\bigr)$
    & $\dot{\omega}_{\min} = \text{min. freq. deriv.},\; \Delta\dot{\omega} = \text{grid spacing}$ \\[6pt]
$n_{\omega}$
    & Index for $\omega_{0} = w_0 + n_{\omega}\,\Delta\omega$
    & $\displaystyle \mathcal{U}\bigl\{\,0,\,N_{\omega}-1\bigr\}$
    & $N_{\omega} = \lfloor(\omega_{\max}-\omega_{\min})/\Delta\omega\rfloor$ \\[6pt]
$n_{\dot{\omega}}$
    & Index for $\dot{\omega}_{0} = \dot{w}_0 + n_{\dot{\omega}}\,\Delta\dot{\omega}$
    & $\displaystyle \mathcal{U}\bigl\{\,0,\,N_{\dot{\omega}}-1\bigr\}$
    & $N_{\dot{\omega}} = \lfloor(\dot{\omega}_{\max}-\dot{\omega}_{\min})/\Delta\dot{\omega}\rfloor$ \\[6pt]
$a$
    & Chirp cosine coefficient ($a = A\cos\phi$)
    & $\displaystyle \mathcal{N}\bigl(0,\,\sigma_{y}^{2}\bigr)$
    & $\sigma_{y} = \mathrm{std}(y)$ \\[6pt]
$b$
    & Chirp sine coefficient ($b = A\sin\phi$)
    & $\displaystyle \mathcal{N}\bigl(0,\,\sigma_{y}^{2}\bigr)$
    & $\sigma_{y} = \mathrm{std}(y)$ \\[6pt]
$m_{i}$
    & Mean flux
    & $\displaystyle \mathcal{N}\bigl(\mu_{y},\,\sigma_{y}^{2}\bigr)$
    & $\mu_{y} = \mathrm{mean}(y),\; \sigma_{y} = \mathrm{std}(y)$ \\
\hline
\end{tabular}
\end{table}

\subsection{Covariance parameters}
\label{appendix:stoch}

For each quasar, we assume known magnitudes of photometric errors given by the same set of $\sigma_{n,k}$'s used to generate the synthetic lightcurve.
Before running the inference, we generate these by drawing $\log \sigma_{n,k} \sim \mathcal{N}(\log(0.2), 0.01)$,
where $\mathcal{N}(x, y)$ is a normal distribution with mean $x$ and variance $y$.
Assuming $N$ observations with timestamps $t_k$, this defines a $N \times N$ diagonal covariance matrix with entries $\sigma_{n,k}^2$, that we take as known and fixed during sampling.

The covariance representing the uncertainty from photometric errors is added to the diagonal of the DRW covariance matrix representing the stochasticity intrinsic to the quasar.
This matrix is constructed at each sampling step contingent on a choice of DRW parameters $\sigma$ and $\tau$, per Eq.~\eqref{eq:drw}.
We draw  $\sigma$ and $\tau$ respectively from a log-normal and a uniform prior distributions, following
\begin{align}
    \log \sigma \sim \mathcal{N}(\log \sigma_y, 4) \, ,\\
    \log \tau \sim \mathcal{U}(\Delta t, T) \, ,
\end{align}
where $\mathcal{U}(x, y)$ is a continuous uniform distribution between $x$ and $y$, $\sigma_y = \mathrm{std}(y)$ is the standard deviation of the observed lightcurve, $\Delta t$ is the median separation between timestamps (${\sim}6$ d) and $T$ is the length of the observation period (10 yr).
We draw $\sigma$ and $\tau$ at each HMC step, together with auxiliary quantities that we use to sample $f_0$ and $\dot f_0$, as we describe next.

\subsection{Nonlinear parameters and the frequency comb}
\label{appendix:comb}

For a given choice of $\sigma$ and $\tau$, the conditional likelihood of any $\{f_0, \dot f_0, A, \phi\}$ is given by Eq.~\eqref{eq:data-generation}.
However, this likelihood is oscillatory as a function of $f_0$ and $\dot f_0$ (or, equivalently, $\omega_0 = 2\pi f_0$ and $\dot\omega_0 = 2\pi\dot f_0$), presenting several false maxima (sidebands) around the true peak.

The sidebands occur at semi-periodic intervals with separation close to $\Delta
\omega_0 \approx 7.72525/T$ in $f$ and $\Delta \dot\omega_0 \approx 15.121/T^2$.
To understand this spacing, first consider the log likelihood for a periodic
signal with angular frequency $\omega$ fit by a model with varying angular
frequency $\omega + \Delta \omega$ in white noise observed for a time $T$.  The
log likelihood $\log\mathcal{L}$ is then proportional to 
\begin{equation}
    \log \mathcal{L}_\omega \propto -\frac{1}{T} \int_0^T \mathrm{d} t \, \left[ \cos \left( \omega t \right) - \cos \left( (\omega + \Delta \omega) t \right) \right]^2 \simeq \frac{\sin\left( \Delta \omega T \right)}{\Delta \omega T} - 1,
\end{equation}
where we have retained only the leading order term for small $\Delta \omega$.  This has a global maximum at $\Delta \omega = 0$ (naturally), and an infinite sequence of local maxima.  The first such secondary local maximum occurs at $\Delta \omega \approx 7.72525/T$, which is the grid spacing we use in our analysis.

Similarly, considering a pure chirp signal with frequency derivative $\dot{\omega}$ fit by a signal with varying frequency derivative $\dot{\omega} + \Delta \dot{\omega}$, the log likelihood is proportional to
\begin{equation}
    \log \mathcal{L}_{\dot{\omega}} \propto -\frac{1}{T} \int_0^T \mathrm{d} t \, \left[ \cos \left( \dot{\omega} t^2/2 \right) - \cos \left( (\dot{\omega} + \Delta \dot{\omega}) t^2/2 \right) \right]^2 \simeq \frac{C\left( T \sqrt{\frac{\Delta \dot{\omega}}{2}} \right)}{T \sqrt{\frac{\Delta \dot{\omega}}{2}}} - 1,
\end{equation}
where we have again kept only the leading term in small $\Delta \dot{\omega}$, and $C(x)$ is the Fresnel cosine integral.  This has a global maximum at $\Delta \dot{\omega} = 0$, and a first local maximum when $\Delta \dot{\omega} \approx 15.121/T^2$, which is the grid spacing we use in our analysis.

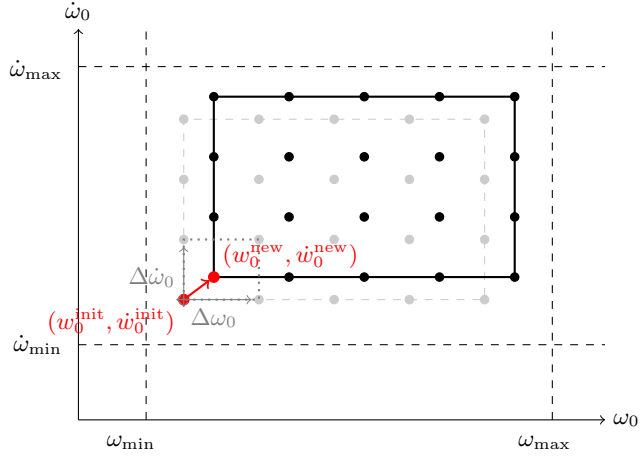
\begin{figure}
    \centering
\begin{tikzpicture}[scale=2.0]
    \def\wmin{0.45}
    \def\wdotmin{0.5}
    \def\dw{0.5}
    \def\dwdot{0.4}
    \def\Nw{5}
    \def\Nwdot{4}

    \draw[->] (0,0) -- (3.5,0) node[right] {$\omega_0$};
    \draw[->] (0,0) -- (0,2.6) node[above] {$\dot\omega_0$};

    \draw[dashed] (\wmin,0) -- (\wmin,2.6);
    \draw[dashed] (0,\wdotmin) -- (3.5,\wdotmin);
    \node at (\wmin-0.1,-0.05) [below] {$\omega_{\rm min}$};
    \node at (-0.05,\wdotmin) [left] {$\dot\omega_{\rm min}$};

    \def\wmidinit{0.7}
    \def\wdotmidinit{0.8}
    \def\wmidnew{0.9}  
    \def\wdotmidnew{0.95}

    \foreach \i in {0,...,\numexpr\Nw-1} {
        \foreach \j in {0,...,\numexpr\Nwdot-1} {
            \filldraw[gray!40] ({\wmidinit + \i*\dw}, {\wdotmidinit + \j*\dwdot}) circle (0.8pt);
        }
    }
    \draw[gray!40, dashed] (\wmidinit,\wdotmidinit) rectangle ++({(\Nw-1)*\dw},{(\Nwdot-1)*\dwdot});

    \foreach \i in {0,...,\numexpr\Nw-1} {
        \foreach \j in {0,...,\numexpr\Nwdot-1} {
            \filldraw[black] ({\wmidnew + \i*\dw}, {\wdotmidnew + \j*\dwdot}) circle (0.8pt);
        }
    }
    \draw[black, thick] (\wmidnew,\wdotmidnew) rectangle ++({(\Nw-1)*\dw},{(\Nwdot-1)*\dwdot});

    \draw[->, red, thick, shorten >=2pt] (\wmidinit, \wdotmidinit) -- (\wmidnew, \wdotmidnew);

    \filldraw[red] (\wmidinit, \wdotmidinit) circle (1pt);
    \node[below left, red] at (\wmidinit, \wdotmidinit) {$(\wmid^{\rm init}, \wdotmid^{\rm init})$};
    \filldraw[red] (\wmidnew, \wdotmidnew) circle (1pt);
    \node[above right, red] at (\wmidnew, \wdotmidnew) {$(\wmid^{\rm new}, \wdotmid^{\rm new})$};

    \draw[<->, gray] (\wmidinit, {\wdotmidinit}) ++(-0.05, 0) -- ++(\dw, 0) node[midway, below, gray] {$\Delta\omega_0$};
    \draw[<->, gray] (\wmidinit, {\wdotmidinit}) ++(0, -0.05) -- ++(0, \dwdot) node[midway, left, gray] {$\Delta\dot\omega_0$};
    \draw[gray, thick, dotted] (\wmidinit, \wdotmidinit) rectangle ++(\dw, \dwdot);

    \draw[dashed] ({\wmidnew + (\Nw-1)*\dw + 0.5*\dw}, 0) -- ({\wmidnew + (\Nw-1)*\dw + 0.5*\dw}, 2.6);
    \draw[dashed] (0, {\wdotmidnew + (\Nwdot-1)*\dwdot + 0.5*\dwdot}) -- (3.5, {\wdotmidnew + (\Nwdot-1)*\dwdot + 0.5*\dwdot});
    \node at ({\wmidnew + (\Nw-1)*\dw + 0.5*\dw - 0.05}, -0.05) [below] {$\omega_{\rm max}$};
    \node at (-0.05, {\wdotmidnew + (\Nwdot-1)*\dwdot + 0.5*\dwdot - 0.02}) [left] {$\dot\omega_{\rm max}$};

\end{tikzpicture}
\caption{Illustration of the frequency comb construction in frequencies and frequency derivative space $(\omega_0, \dot\omega_0)$. The HMC sampler first picks a lower-left corner $(\wmid^{\rm init}, \wdotmid^{\rm init})$ near $(\omega_{\rm min}, \dot\omega_{\rm min})$, following Eq.~\eqref{eq:wmid-wdotmid-draw}, which is then used to construct an $N_{\omega} \times N_{\dot\omega}$ grid  with spacing $\Delta \omega_0$ and $\Delta \dot\omega_0$ (gray dots); this grid becomes the basis for a Gibbs step, which will sample a random location $(n_\omega, n_{\dot\omega})$ within it.
Next, the HMC sampler picks a new lower-left corner $(\wmid^{\rm new}, \wdotmid^{\rm new})$, which is used to construct a new grid  (black dots), and the process repeats.
The dashed lines indicate the minimum and maximum frequencies and frequency derivatives that will be probed during sampling.}
    \label{fig:frequency-comb}
\end{figure}

The local maxima in the likelihood as a function of $\omega$ and $\dot{\omega}$
can cause the HMC sampler to become locally stuck, without finding the true
typical set of the posterior.  To address this, we use our knowledge of the
approximate separation between likelihood modes to construct a \emph{grid} (or
\emph{comb}) of frequencies and frequency derivatives that helps the sampler hop
between potential modes of the likelihood, while also smoothly sliding the grid
to cover the desired frequency space.
We illustrate this in Figure~\ref{fig:frequency-comb}, and describe it below.

Assume that we want to look for frequencies $\omega_0$ within some range $\omega_{\rm min} \leq \omega_0 \leq \omega_{\rm max}$ and frequency derivatives $\dot\omega_0$ within $\dot\omega_{\rm min} \leq \dot\omega_0 \leq \dot\omega_{\rm max}$.
Given the above spacings $\Delta \omega_0$ and $\Delta \dot\omega_0$, the desired $\omega_0$ and $\dot\omega_0$ ranges can be covered by a grid with a number of frequency and frequency-derivative points given by
\begin{subequations}
    \begin{align}
        N_\omega &= \left\lfloor (\omega_{\rm max} - \omega_{\rm min})/\Delta \omega_0\right\rfloor,\\
        N_{\dot\omega} &= \left\lfloor (\dot\omega_{\rm max} - \dot\omega_{\rm min})/\Delta \dot\omega_0\right\rfloor
    \end{align}
\end{subequations}
where $\lfloor \cdot \rfloor$ is the floor function; a location on the grid is specified by indices $(n_\omega, n_{\dot\omega})$ with $0 \leq n_\omega < N_\omega$ and $0 \leq n_{\dot\omega} < N_{\dot\omega}$.

To allow one of the gridpoints to snap onto the bulk of the likelihood, we free the grid to continuously slide slightly up and down in $\omega_0$ and $\dot\omega_0$ space by up to an amount commensurate with the grid spacing (Fig.~\ref{fig:frequency-comb}).
Concretely, letting  $(\wmid, \wdotmid)$ define the location of the lower-left corner of the grid in $(\omega_0, \dot\omega_0)$ space, i.e., the placement of the $(n_\omega=0, n_{\dot\omega}=0)$ grid point, we slide the grid by sampling
\begin{subequations}
    \label{eq:wmid-wdotmid-draw}
    \begin{align}
        \wmid \sim \mathcal{N}_{[0, \infty)}(\omega_{\rm min}, \Delta \omega_0^2)\, , \\
        \wdotmid \sim \mathcal{N}_{[0, \infty)}(\dot\omega_{\rm min}, \Delta \dot\omega_0^2)\, ,
    \end{align}
\end{subequations}%
where $\mathcal{N}_{[0, \infty)}(x, y)$ is a normal distribution with mean $x$ and variance $y$, but truncated to positive values.
That is, at each HMC step, we choose an anchor point $(\wmid, \wdotmid)$ for the grid, which cannot fall below the minimum values of $\omega_{\rm min}$ and $\dot\omega_{\rm min}$ and will typically not be farther from the $(\omega_{\rm min}, \dot\omega_{\rm min})$ corner than the grid spacing $\Delta \omega_0$ and $\Delta \dot\omega_0$.
Based on that anchor point and spacing, we can then construct an $N_\omega \times N_{\dot\omega}$ grid of frequencies and frequency derivatives.

\begin{figure}[htbp]
  \centering
  \begin{subfigure}[c]{0.45\textwidth}
    \centering
    \includegraphics[height=0.7\textwidth]{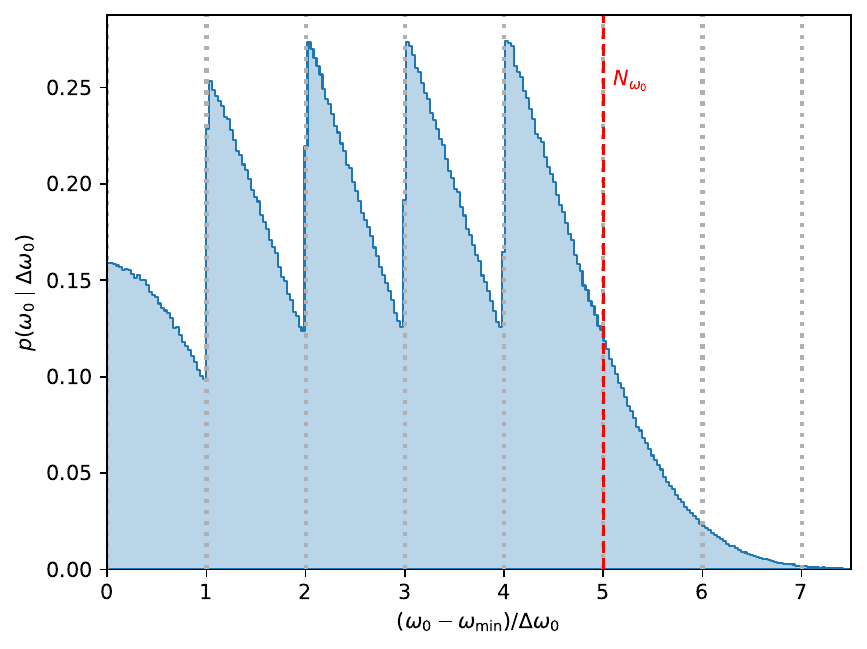}
    \caption{}
    \label{fig:omega-prior}
  \end{subfigure}
  \quad
  \begin{subfigure}[c]{0.45\textwidth}
    \centering
    \includegraphics[height=0.65\textwidth]{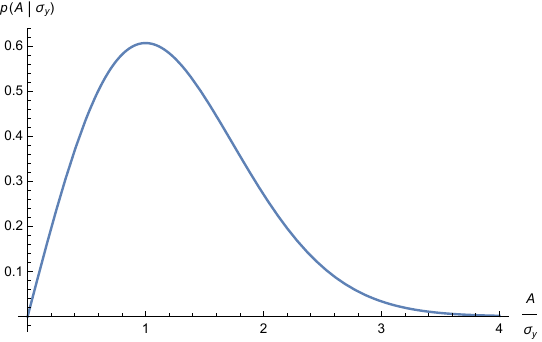}
    \caption{}
    \label{fig:prior-A}
  \end{subfigure}
  \caption{Effective priors for the frequency~$\omega_0$ (left) and chirp amplitude~$A$ (right). (a) The effective prior on $\omega_0$ results from the sampling strategy associated with Eq.~\eqref{eq:omega0-wdot0-prior} which defines a frequency grid of $N_{\omega_0}$ points separated by $\Delta \omega_0$ and starting at values not smaller than $\omega_{\rm min}$ (for the figure, we set $N_{\omega_0} = 5$); the distribution is broad, with spikes at integer multiples of $\Delta \omega_0$.
    (b) The effective prior on $A$ is a Rayleigh distribution with scale $\sigma_y = {\rm std}(y)$, per Eq.~\eqref{eq:a-prior}.
    The prior for $\dot\omega_0$ is analogous to (a), with a grid of $N_{\dot\omega_0}$ points separated by $\Delta \dot\omega_0$ and starting at values not smaller than $\dot\omega_{\rm min}$.
    }
  \label{fig:combined-priors}
\end{figure}

Finally, we use a Gibbs step to select a location on the grid $(n_\omega,
n_{\dot\omega})$ by sampling from a categorical distribution with a prior such
that
\begin{subequations}
    \begin{align}
        n_\omega &\sim \mathcal{U}\{0, N_\omega - 1\}\, , \\
        n_{\dot\omega} &\sim \mathcal{U}\{0, N_{\dot\omega} - 1\}\, ,
    \end{align}
\end{subequations}%
where $\mathcal{U}\{j, k\}$ is a discrete uniform distribution over the integers $j$ to $k$.
With such a choice, we construct values of the frequency $\omega_0$ and frequency derivative $\dot\omega_0$ at the grid point as
\begin{subequations} \label{eq:omega0-wdot0-prior}
    \begin{align}
        \omega_0 &= \wmid + n_\omega \Delta \omega_0\, , \\
        \dot\omega_0 &= \wdotmid + n_{\dot\omega} \Delta \dot\omega_0\, .
    \end{align}
\end{subequations}%
The effective prior on $\omega_0$ and $\dot\omega_0$ is sufficiently broad to cover the desired range (shown in Fig.~\ref{fig:omega-prior} for $\omega_0$, analogous for $\dot\omega_0$).
This is all we need to compute the likelihood at this grid point, marginalizing over $A$ and $\phi$ as described below.

\subsection{Linear parameters and analytic marginalization}
\label{appendix:quadratures}

Having chosen $\omega_0$, $\dot\omega_0$, $\sigma$ and $\tau$, we could also sample the chirp amplitude $A$, phase $\phi$, and mean $m_i$ to directly compute the likelihood of Eq.~\eqref{eq:data-generation};
however, this would be suboptimal.
Instead, we take advantage of the Gaussian form of Eq.~\eqref{eq:data-generation} to analytically marginalize over $\{A, \phi, m_i\}$ for any given choice of $\{\omega_0, \dot\omega_0, \sigma, \tau\}$, thus effectively reducing the number of parameters to sample from 7 to 4.
This vastly improves sampling efficiency without losing information about the suppressed parameters, which can be restored in post-processing.

The analytical marginalization is made possible by factorization properties of the Gaussian \citep{Hogg:2020jwh}, which we can fully leverage by reparameterizing the chirp template in Eq.~\eqref{eq:chirp}.
Concretely, we re-express the chirp amplitude $A$ and phase $\phi$ in terms of the corresponding \emph{linear quadratures} \citep[e.g.,][]{Isi:2022mbx},
\begin{align}
    a = A \cos \phi ~~, ~~ 
    b = A \sin \phi , 
\end{align}
so that $A = \sqrt{a^2 + b^2}$ and $\phi = \tan^{-1}(b/a)$.
In terms of these quantities, the template of Eq.~\eqref{eq:chirp} becomes
\begin{subequations}
\begin{align}
    y_{\rm chirp}(t) &= a \cos [2 \pi f(t)] + b \sin[2 \pi f(t)]\, , \\
    &= a\, y_c(t) + b\, y_s(t) \, ,
\end{align}
\end{subequations}%
with $y_c(t; \omega_0, \dot\omega_0) = \cos[2 \pi f(t)]$ and $y_s(t) = \sin[2 \pi f(t)]$ implicit functions of $\omega_0$ and $\dot\omega_0$.
Defining also $y_m(t) = 1$, we can write the full template as
\begin{align}
    m_i + y_{\rm chirp}(t; a, b) &= m_i\, y_m(t) + a\, y_c(t) + b\, y_s(t)\, .
\end{align}
This reparameterization is advantageous because the $a$, $b$ and $m_i$ parameters enter the waveform as linear coefficients to produce combinations of the basis functions $y_c(t)$, $y_s(t)$ and $y_m(t)$, which will allow us to analytically marginalize out $\{m_i, a, b\}$ conditioned on $\{\omega_0, \dot\omega_0, \sigma, \tau\}$.
To affect the analytical marginalization, we choose Gaussian priors on $a$, $b$ and $m_i$ such that
\begin{subequations}
    \label{eq:prior-a-b-m}
\begin{align}
    a &\sim \mathcal{N}(0, \sigma_y^2)\, , \\
    b &\sim \mathcal{N}(0, \sigma_y^2)\, , \\
    m_i &\sim \mathcal{N}(\mu_y, \sigma_y^2)\, ,
\end{align}
\end{subequations}%
where, as above, $\sigma_y = \mathrm{std}(y)$ is the standard deviation of the observed lightcurve and $\mu_y = \mathrm{mean}(y)$ is its mean.
With this choice of prior $p(a, b, m_i)$, we can compute the marginalized likelihood
\begin{equation}
\label{eq:marginalized-likelihood}
    p(y \mid \omega_0, \dot\omega_0, \sigma, \tau) = \int p(y \mid  a, b, m_i, \omega_0, \dot\omega_0, \sigma, \tau)\, p(a, b, m_i)\, \mathrm{d}a\, \mathrm{d}b\, \mathrm{d}m_i\, ,
\end{equation}
analytically as explained in \citet{Hogg:2020jwh}, i.e., without carrying out the integral explicitly.

Using this framework, the HMC-Gibbs sampler produces a set of samples in $\{\omega_0, \dot\omega_0, \sigma, \tau\}$, drawn from the marginal posterior
\begin{equation}
    p(\omega_0, \dot\omega_0, \sigma, \tau \mid y) = \int p(\omega_0, \dot\omega_0, \sigma, \tau, a, b, m_i \mid y)\, \mathrm{d}a\, \mathrm{d}b\, \mathrm{d}m_i\, .
\end{equation}
We restore the ${a, b, m_i}$ parameters in post-processing by sampling from the conditional posterior
$p(a, b, m_i \mid \omega_0, \dot\omega_0, \sigma, \tau, y)$, which is a Gaussian that can be computed analytically 
for each sampled point in $\{\omega_0, \dot\omega_0, \sigma, \tau\}$ \citep{Hogg:2020jwh}.
This uses the fact that
\begin{equation}
    p(\omega_0, \dot\omega_0, \sigma, \tau, a, b, m_i \mid y) = p(a, b, m_i \mid \omega_0, \dot\omega_0, \sigma, \tau, y)\, p(a, b, m_i \mid y) / p(\omega_0, \dot\omega_0, \sigma, \tau \mid y)
\end{equation}
by Bayes theorem.

The analytical marginalization requires setting a Gaussian prior on the linear parameters $a$, $b$ and $m_i$, as in Eq.~\eqref{eq:prior-a-b-m}.
It can be shown that this is equivalent to a uniform prior on $\phi$, but not on $A$ \citep[e.g.,][]{Isi:2022mbx};
rather, the induced prior on $A$ is a Rayleigh distribution with scale $\sigma_y$, i.e.,
\begin{equation} \label{eq:a-prior}
    p(A \mid \sigma_y ) = \frac{A}{\sigma_y^2} \exp\left(-\frac{A^2}{2\sigma_y^2}\right) ,
\end{equation}
as we illustrate in Figure~\ref{fig:prior-A}.
Although this distribution is not flat, we find that it is sufficiently broad to not bias our inference.

\section{Inference of the chirp} \label{appendix:snr_fdot}

\begin{figure*}
    \centering
    \hspace{-10mm}
    \includegraphics[width=0.54\columnwidth]{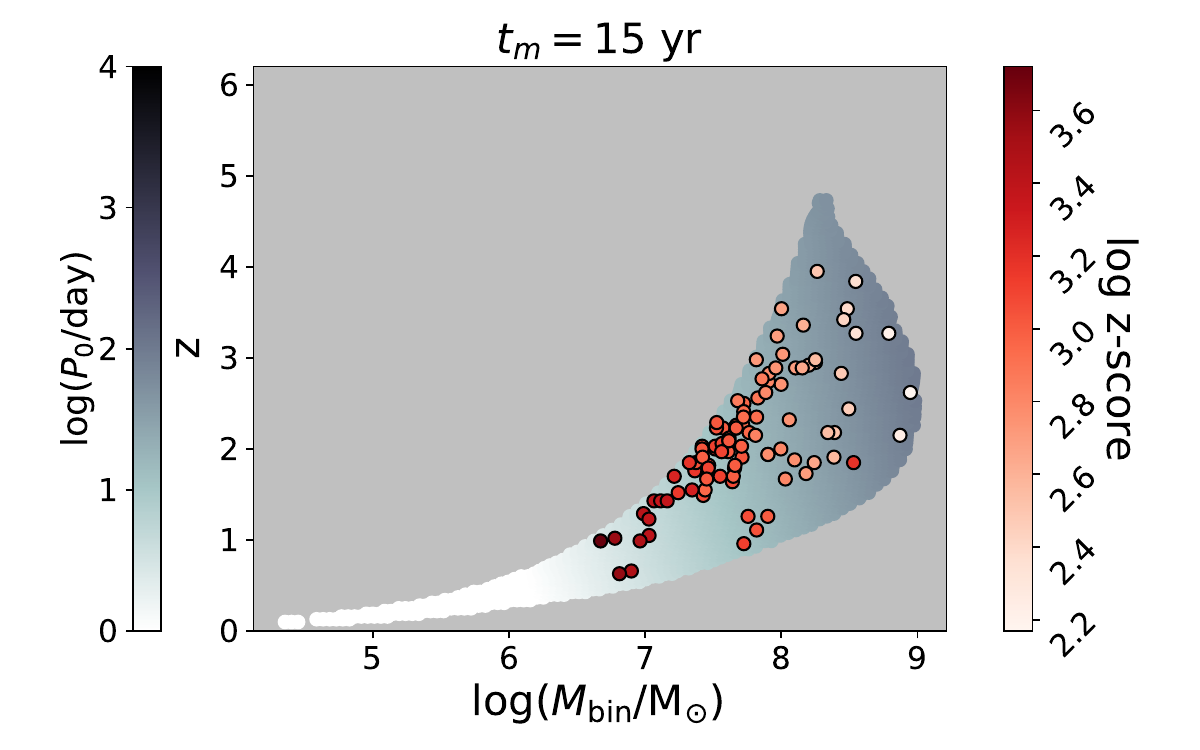} \hspace{-8mm}
    \includegraphics[width=0.54\columnwidth]{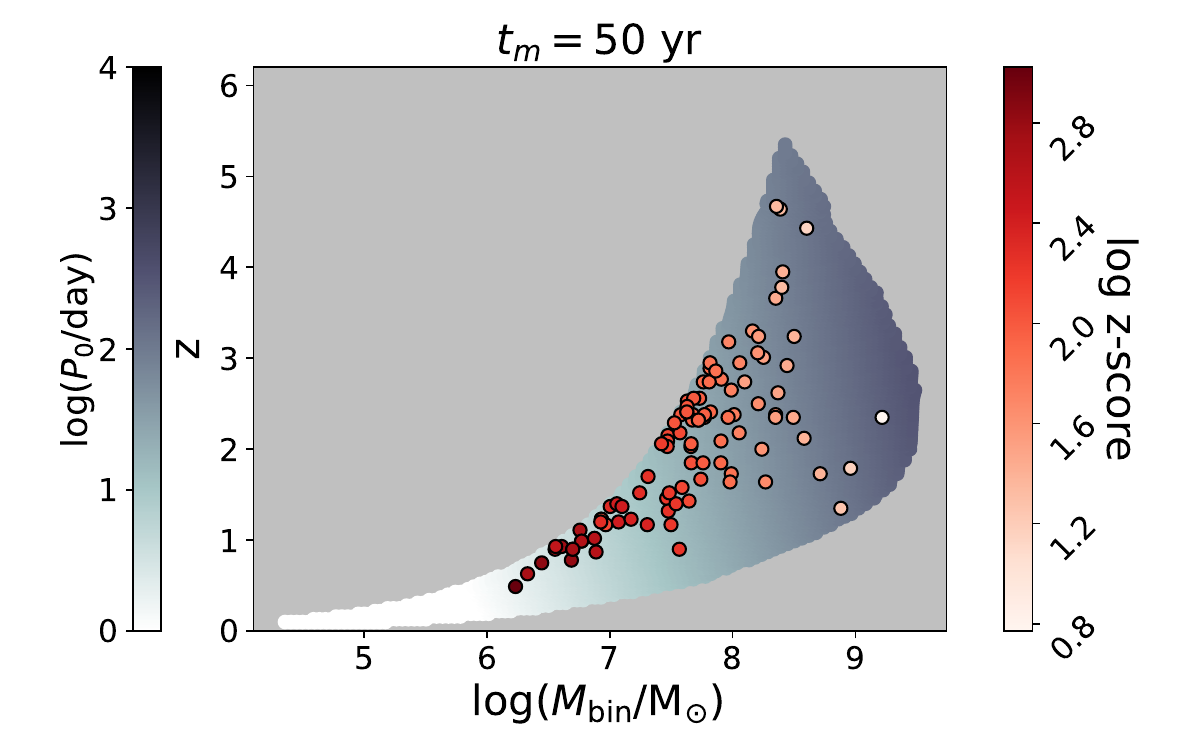} \\
    \hspace{-10mm}
    \includegraphics[width=0.54\columnwidth]{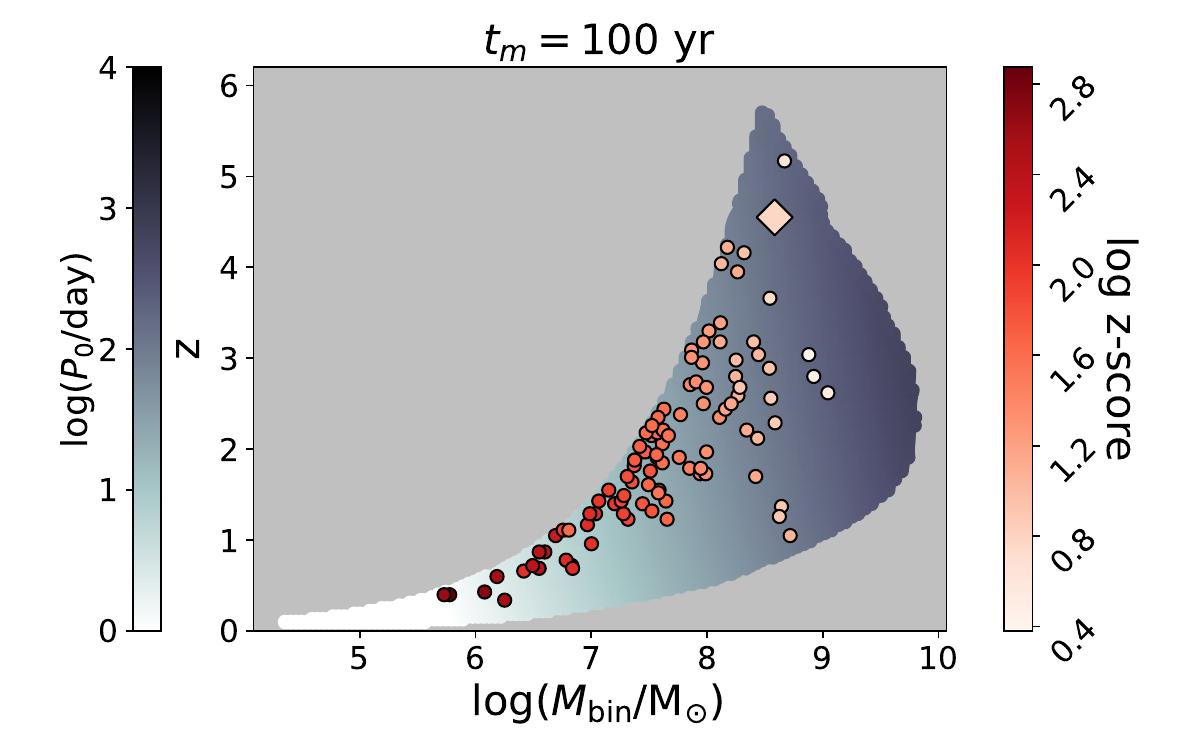} \hspace{-8mm}
    \includegraphics[width=0.54\columnwidth]{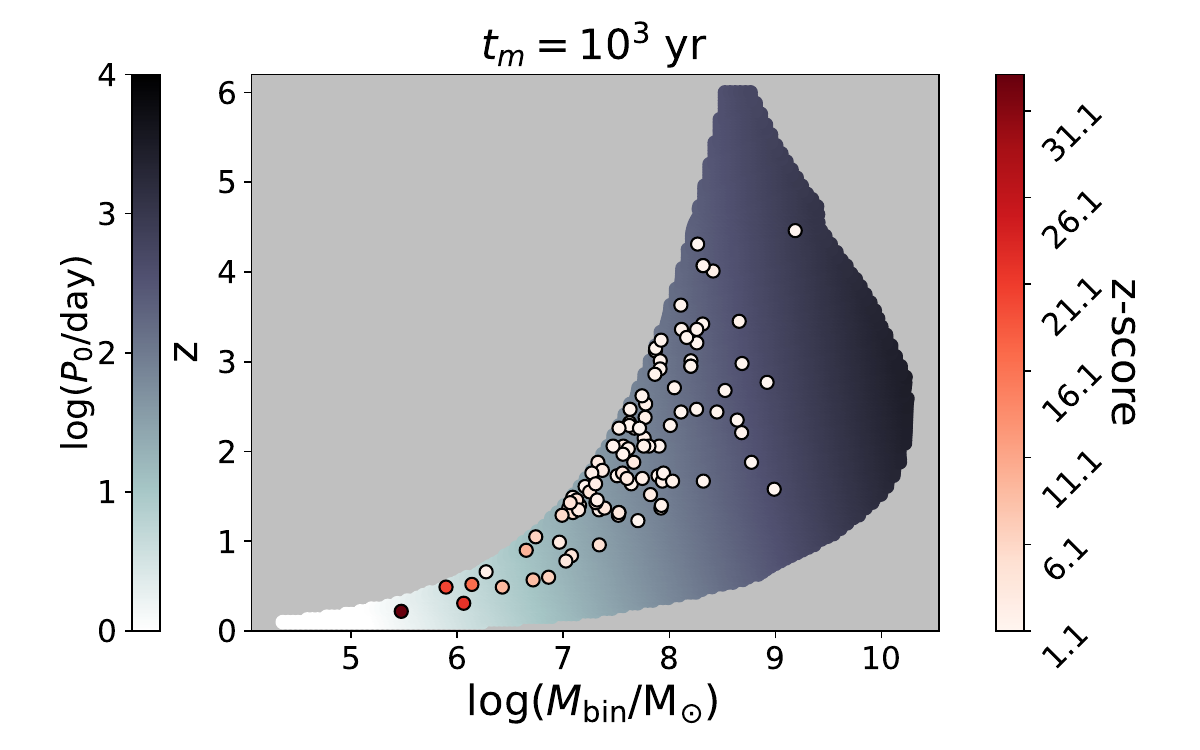} \\
    \includegraphics[width=0.54\columnwidth]{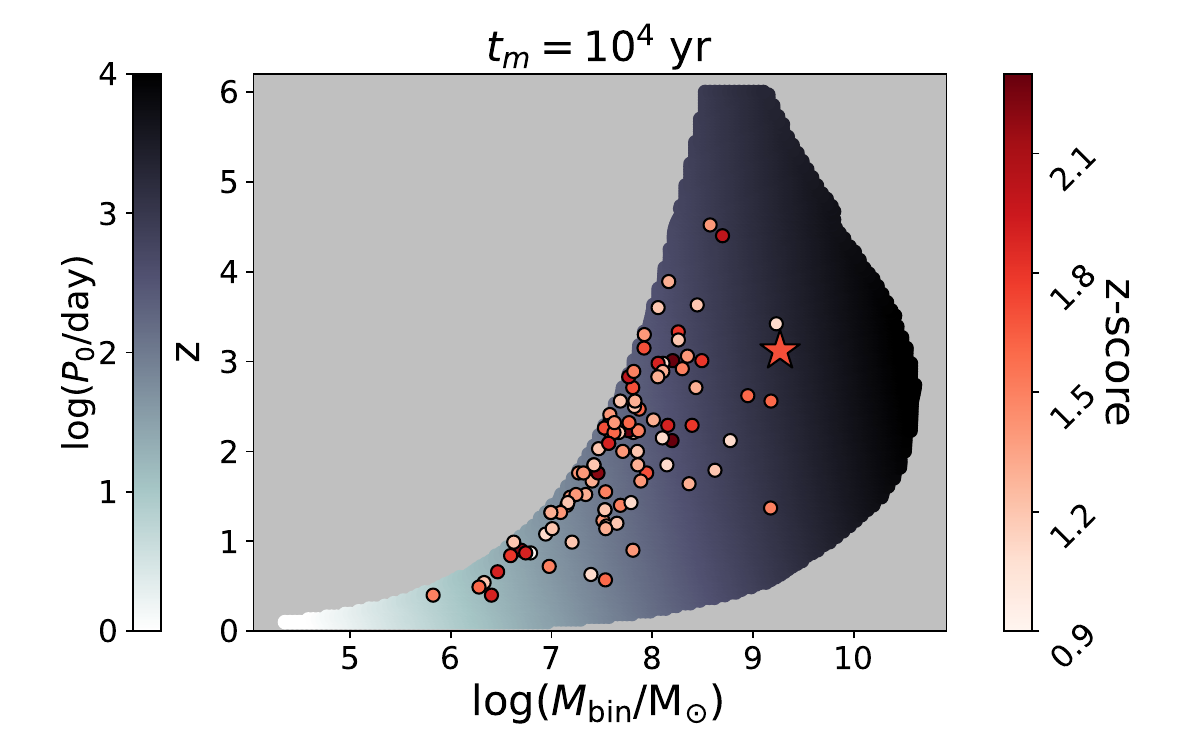}
    \caption{We show the chirp ($\dot f_0$) posterior z-score's for the same binaries presented in Fig.~\ref{fig:achirp_p0}. Notice the difference in ranges and scales on the z-score  (red) color gradients. The diamond and the star are the two wide binaries that we highlight in \S~\ref{subsec:low-high-chirp}. 
    }
    \label{fig:fdot0_p0}
\end{figure*}

We demonstrate our ability to measure chirp as a function of binary time to merger, total mass and orbital period in Fig.~\ref{fig:fdot0_p0}. The z-score of $\dot f_0$ drastically drops for longer orbital periods (and consequently for larger $t_m$), further emphasizing on the importance of observing many cycles to achieve reliable measurement of chirp. 

\section{Runtime} 
\label{appendix:computational-cost}

\begin{figure}
        \centering
        \includegraphics[width=0.68\textwidth]{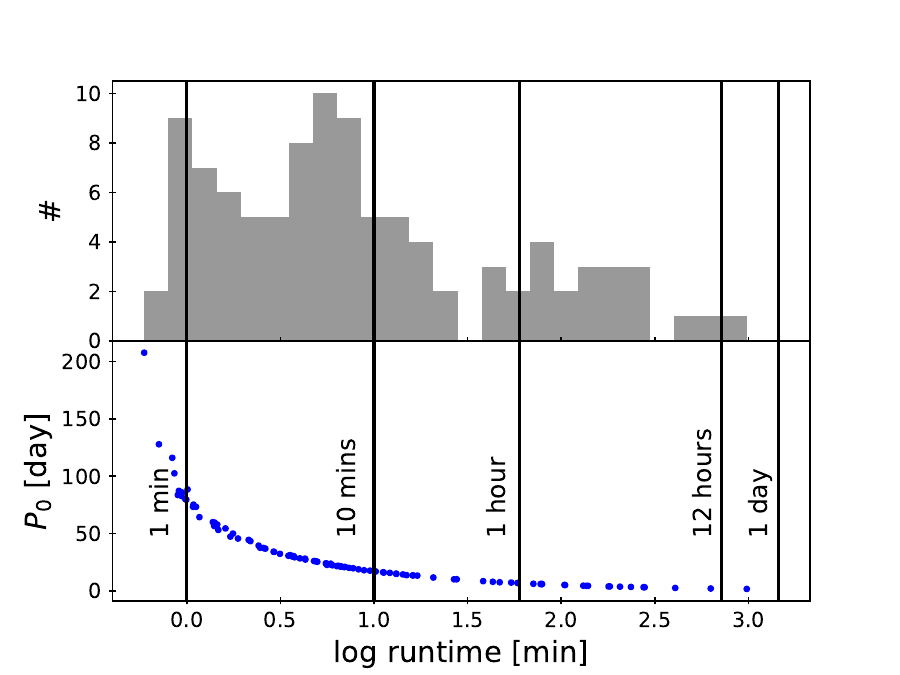}
        \caption{The runtime of our Bayesian inference model for each lightcurve in the fiducial binary population (top right panel in Figure~\ref{fig:achirp_p0}). The log-runtime are presented in the histogram (top panel), which shows that most of our lightcurves are analyzed within 10 minutes. We also show runtime versus orbital periods (bottom panel), and we find that the runtime are shorter for binaries with longer periods. }
        \label{fig:runtime}
    \end{figure}

Ultimately, we will apply our analysis to systematically search for compact MBH binaries in the LSST, beginning in 2025 - 2026. In Figure~\ref{fig:runtime}, we show the runtime of our Bayesian algorithm on the fiducial 100 binaries with $t_m=50$ yr. Most lightcurves are analyzed under 10 minutes, with maximum runtime of $\sim1$ day. This is significantly more efficient than other Bayesian algorithms developed to identify periodicity in quasar lightcurves, which typically take at least a few hours per lightcurve \citep[e.g.][]{Witt2022}. 
With our current average runtime, we can perform our analysis on 100 million quasars using $\sim 16$ million CPU hours. In future work, we will work on optimizing the algorithm to speed up the search on big datasets. 
Additionally, the search space for binary parameters, including periodicity and chirp, are much wider than what we consider in this work, which will make the runtime longer.
On the other hand, lightcurves with ``obvious" periodicity can be identified with certain `pre-selection' (faster) approach, like the LS periodogram, to narrow down the parameter space for the Bayesian analysis. Other parameters can undergo similar pre-search steps using simpler statistics. We will design the pre-selection methods for each parameter for a given  observation in future work.

\label{subsec:real_lsst}

\bibliographystyle{aasjournal}
\bibliography{cx}

\end{document}